\documentclass{aa}  
\usepackage{float}
\usepackage{lscape}
\usepackage{longtable}
\usepackage{geometry}
\usepackage{caption}
\usepackage{booktabs}
\usepackage{graphicx}
\usepackage[normalem]{ulem}
\usepackage{txfonts}
\usepackage[colorlinks=true, linkcolor=blue]{hyperref}
\hypersetup{
  colorlinks = true,
  linkcolor  = blue,
  citecolor  = blue,
  urlcolor   = blue
}
\usepackage[utf8]{inputenc}

\newcommand{\teff}{T_\mathrm{eff}}

\begin{document} 

\title{Exploring stellar activity in a sample of active M dwarfs}
\authorrunning{Rajpurohit, A. S.}

\author{Rajpurohit, A. S. \inst{1}, Kumar, V. \inst{2}, Mudit, K. Srivastava \inst{1}, Labadie, L. \inst{2}, Rajpurohit, K. \inst{3}, Fern\'andez-Trincado, J. G. \inst{4}}
\institute{Astronomy \& Astrophysics Division, Physical Research Laboratory, Navrangapura, Ahmedabad 380009, India \\
\email{arvindr@prl.res.in}
\and
I. Physikalisches Institut, Universit\"at zu K\"oln, Z\"ulpicher Stra\ss e 77, 50937, K\"oln, Germany
\and
Center for Astrophysics $|$ Harvard \& Smithsonian, 60 Garden   Street, Cambridge, MA 02138, USA
\and
Instituto de Astronom\'ia, Universidad Cat\'olica del Norte, Av. Angamos 0610, Antofagasta, Chile, \email{jose.fernandez@ucn.cl}
}
   
\date{Received XXX; accepted XXX}

 
\abstract
{
Most M dwarfs present high chromospheric activity that can exceed the solar magnetic activity. This can substantially influence planetary, atmospheric, and biological processes, impacting the habitability of orbiting planets. Therefore, characterizing the magnetic activity of M dwarfs is very important for understanding the physical mechanisms responsible for it, which are the primary targets in the search for exoplanets within the habitable zone.
}
{
This study aims to characterize the stellar activity of active M dwarfs by understanding the relations between magnetic activities, stellar parameters, and flare properties.
}
{
We analyzed TESS photometric data combined with spectroscopic observations of active M dwarfs. We examined the relationship between the flare occurrence rate, flare energies, rotation period, filling factor, and chromospheric activity indicators. Furthermore, the correlation between flare amplitude and duration and cumulative flare energy frequency distributions was investigated to probe the underlying mechanisms driving magnetic activity in flaring M dwarfs.
}
{
We find that the flare occurrence rate displays a flat distribution across spectral types M0--M4 ($T_{\mathrm{eff}} \sim 3900\text{--}3200\,\mathrm{K}$) and declines for later types. Faster rotators with $P_{\rm rot} < 1$ day exhibit a higher flare occurrence rate and flare activity. M dwarfs with a higher flare occurrence rate tend to exhibit lower flare amplitudes, indicating that frequent flares in these M dwarfs are generally less energetic. Within the mass range of $0.15$--$0.76\,M_{\odot}$, the median of $L_{\mathrm{H\alpha}} / L_{\mathrm{bol}}$ in evenly divided mass bins of $\sim 0.1\,M_{\odot}$ varies by a factor of $\sim 2.5$, while $\Delta$EW decreases by $92\%$ across the sample. We derive power-law indexes of the cumulative flare frequency distribution for M dwarf subgroups, indicating a decreasing trend from M0 to M5 dwarfs with a value of $\alpha$ from $1.68$ to $1.95$, respectively.
}
 {
 We characterized the stellar activity in M dwarfs through chromospheric indicators like H$\alpha$ emission, star-spot coverage, flare occurrence rates, flare energies, and flare duration. Our results suggest a stellar activity transition near M4, with stronger H$\alpha$ emission linked to higher flare occurrence. Rapid rotators ($P_{\mathrm{rot}}$ < 1 day) exhibit significantly higher flare occurrence rate, supporting the idea that strong magnetic dynamos in fast-rotating M dwarfs sustain frequent flaring activity. Our analysis confirms that highly active stars dissipate magnetic energy through numerous low-energy flares rather than fewer high-energy events. We also show that chromospheric activity and flare activity follow a power-law relationship.
 }

\keywords{techniques: spectroscopic-- stars: activity -- stars: late-type -- stars: flare -- stars: rotation}

\maketitle

\section{Introduction}

M dwarfs are the coolest stars with $\teff$ ranging from 2500–4000 K and the least massive (0.075–0.6$M_{\odot}$) stellar objects in the Milky-Way Galaxy. They contribute roughly 40$\%$ of the total stellar mass \citep{Henry1998} and account for about 70$\%$ of the stellar population \citep{Reid1995, Reyle2021}. M dwarf populations show a broad range of evolutionary stages and chemical compositions, with young, metal-rich M dwarfs in open clusters, while older, metal-poor M dwarfs can be found in the Galactic halo \citep{Green1994}. M dwarfs have gained particular interest in the search for habitable exoplanets. Studies by \cite{Luque2022, Kossakowski2023} and \cite{Donati2023, Donati2024} have unveiled the presence of exoplanets orbiting M dwarfs. \cite{Klein2022} discovered exoplanets around the bright, young active M-dwarf AU Mic, whereas \cite{Anglada2016} revealed the presence of a terrestrial planet around the nearest to M dwarfs to the Sun, Proxima Centauri.

Various optical and near-infrared (NIR) observational facilities such as HARPS \citep{Mayor2003}, HARPS-N \citep{Cosentino2012}, CRIRES \citep{Kaeufl2004}, CARMENES \citep{Quirrenbach2014}, SPIRou \citep{Cersullo2017}, HPF \citep{Mahadevan2012}, and NIRPS \citep{Bouchy2025} have started yielding high-quality, high signal-to-noise ratio spectra. Such high-quality spectra offer a powerful tool to investigate the properties of M dwarfs, such as their fundamental parameters and radial velocities, which enable the detection of low-mass exoplanets \citep{Astudillo2017, Amado2021}. Further, it allows us to determine their elemental abundances, line broadening, and Zeeman splitting in magnetically sensitive lines, which offer constraints on their rotation.

The active nature of M dwarfs substantially influences the atmosphere of planets orbiting them because of the high-energy radiation they receive, thus impacting their potential habitability \citep{Tilley2019}. Compared to hotter stars like our Sun, M dwarfs are known to be more active. \cite{West2004}  suggested that the fraction of active M dwarfs increases from M0 to M8 and then in later spectral types, particularly in the brown dwarfs (L3-L4) regime. The magnetic activity can generate variability in radial velocity (RV) measurements, often mimicking periodic signals akin to those caused by actual exoplanets. Thus, to assess the habitability of exoplanets orbiting these stars, the evolution of planetary systems in general, and to explore various physical mechanisms underlying these signals generated by stellar activity, it is necessary to understand the magnetic activity of M dwarfs.

Spectroscopic and photometric tools have been used quite extensively to study stellar activity in M Dwarfs. Various spectral lines such as \text{Mg\,\textsc{ii}}, \text{Ca\,\textsc{ii}}, \text{Na\,\textsc{i}}, \text{K\,\textsc{i}}, and H$\alpha$ are widely used to assess the chromospheric activity within the optical spectral range, covering the temperature minimum region up to the upper chromosphere \citep{Cincunegui2007, Diaz2007b, Cao2014}. The variations in the mean line-core flux of these lines can then be used to indicate the overall activity level of M Dwarfs. Flux variations in these lines provide valuable insights into the different regions within the stellar atmosphere at varying heights. For instance, the H$\alpha$ emission line forms in the upper chromosphere, \text{Na\,\textsc{i}} lines originate in the lower chromosphere, and \text{Ca\,\textsc{ii}} and \text{K\,\textsc{i}} lines span from the middle chromosphere to the upper chromosphere \citep{Mauas1994, Leenaarts2012, Fontenla2016}.\\

One method of probing the stellar activity is to investigate flares and use them as a proxy for magnetic activity. Flares are short-lived yet powerful energetic phenomena that take place on main-sequence stars, where magnetic energy is converted into transient emissions spanning a broad spectrum of wavelengths, from radio to X-ray. These events also drive plasma heating, particle acceleration, and large-scale plasma motion. \citep{Benz2010, Crowley2024}.  \cite{Shibayama2013} suggested that stellar flares follow power law relation $\frac{dN}{dE} \propto E^{-\alpha}$ with $\alpha \sim 2$. These flares are believed to result from a magnetic reconnection event, which produces a beam of charged particles that collide with the stellar photosphere. This interaction triggers intense heating and emits radiation across nearly the entire electromagnetic spectrum \citep{Davenport2016}. It is believed that during the flare, magnetic reconnection caused a change in field line topology, leading to a significant energy release, often on the order of magnitude of X-class solar flares, or even greater \citep{Pettersen1989, Doyle2018}. Also, the time scales of flares in stars are unpredictable. It can range from a few hours to a few days. Thus, obtaining a systematic sample of flares for an individual star has been highly resource-demanding and is done for only a few particularly active stars.   

Low mass stars, particularly M dwarfs with deepest convective zones, flare more frequently than G or K dwarfs \citep{Walkowicz2011}. In mid to late M dwarfs, the rate of occurrence of flares is nearly $30\%$, whereas, in early M dwarfs, it is $5\%$ and less than $1\%$ of stars with F, G, and K spectral types \citep{Gunther2020}. During flaring, the amount of energy released ranges from 10$^{29}$ to 10$^{32}$  erg \citep{Parnell2000, Shibayama2013}. It is found that fast-rotating M dwarfs flare more frequently. Such fast-rotating M dwarfs with rotation period ($P_{\mathrm{rot}}$) $<$ 10 days show significant variations in their light curves, possibly due to star-spots, which are believed to be imprints of magnetic field lines on their photosphere. However, studies by \cite{Ramsay2013}, \cite{Davenport2014}, and \cite{Doyle2018} have presented evidence challenging this view. They observed no correlation between the number of flares, their amplitudes, and the rotational phase in M dwarfs studied with Kepler and K2. This suggests that flares may occur independently of a large, dominant star-spot. Also, in slow-rotating M dwarfs, the ratio of X-rays, H$\alpha$, and Ca\,H\&K flux to bolometric luminosity declines rapidly, whereas the activity remains saturated for fast rotators \citep{Raetz2020}. Also, in fast-rotating M dwarfs, large spots may exist and probably are located randomly on the surface at a very small length scale \citep{Magaudda2020}.

Photometric surveys, including Kepler, its successor K2 \citep{Howell2014}, and the Transiting Exoplanet Survey Satellite (TESS) \citep{Ricker2015}, have significantly increased interest in studying magnetic phenomena in host stars due to their impact on exoplanets. These surveys observe large samples with high photometric precision, enabling the detection of magnetic activity indicators such as flares, periodic luminosity variations \citep{Reinhold2017}, and star-spot variability, which helps identify stellar activity cycles. Using the high-precision photometry from the Kepler survey \cite{Maehara2012}, we found many energetic superflares, which are even 10$^{4}$ times larger than solar flares. Using such high-precision photometric data, strong correlations between the flare energy, amplitude, duration, and decay time have been found by \cite{Hawley2014, Raetz2020}, and \cite{Yang2023}. Also, the contrast between flares and a star's quiescent state is more prominent in M dwarfs, leading to fewer studies on flares in field G dwarfs. For inactive stars like our Sun, the flare rates are highly unconstrained. 

The primary objective of this study is to investigate stellar flares and magnetic activity in M dwarfs through combined spectroscopic and photometric observations. In particular, we examine the relationship between flare properties such as flare energy $\langle E_{\mathrm{flare}} \rangle$, flare frequency, and flare occurrence rate (FOR) and various stellar activity indicators, including spectral type, $P_{\mathrm{rot}}$, flare frequency distribution (FFD), and star-spot-filling factor ($f_{\mathrm{s}}$). Section 2 presents the target sample along with spectroscopic and photometric observations. Section 3 describes the methods used for chromospheric activity analysis and flare parameters. The results and their implications are discussed in Section 4, and our conclusions are in Section 5.

\section{Data}
\subsection{Target sample}
\begin{figure}[hbt!]
	\includegraphics[angle=0,width=0.48\textwidth]{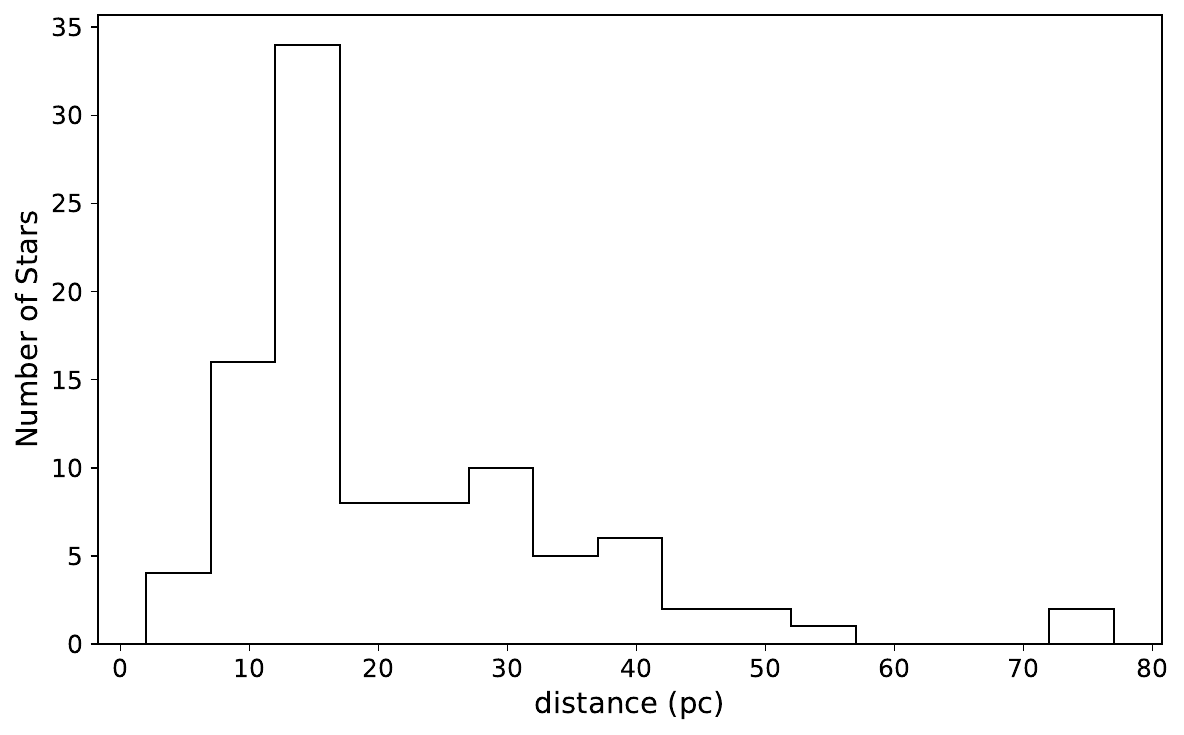}
    \includegraphics[angle=0,width=0.48\textwidth]{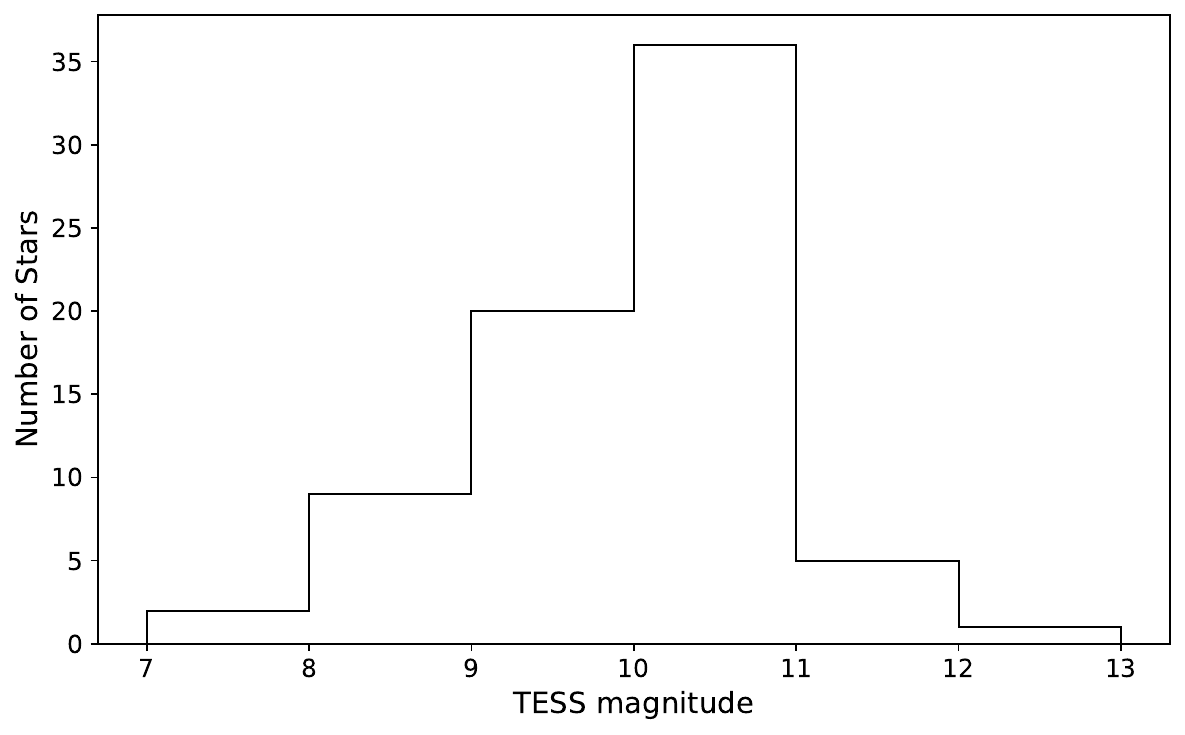}
	\caption{Distance distribution (top panel) of stars in our sample, binned at 5 pc intervals and TESS magnitude distribution (bottom panel) of the same sample.}
	\label{fig1}
\end{figure}

In this current study, our sample consists of bright field active M dwarfs with spectral types ranging from  M0 to M8.5, selected from the catalog of \citet{Kumar2023}. The majority of M dwarf targets in our sample lie on the main sequence, with only a small fraction showing pre-main-sequence ages based on the StarHorse ages \citep{Queiroz2018, Anders2019} reported in \citet{Kumar2023}. The targets in our sample lie at intermediate to high Galactic latitudes (typically $|b| \gtrsim 15\!-\!20^{\circ}$), well away from the Galactic plane. Hence, it minimizes the likelihood of contamination from young stellar associations. Fig.~\ref {fig1} shows the distance distribution and the TESS magnitude distribution of all the stars in our sample. The distances and TESS magnitudes are from \cite{Stassun2018}.

\subsection{Spectroscopic activity parameters}
In our previous work \citep{Kumar2023}, we conducted low-resolution (R$\sim$1000) spectroscopic monitoring of active M dwarfs (M0–M6.5) using Mount-Abu Faint Object Spectrograph and Camera- Pathfinder (MFOSC-P) \citep{Srivastava2018, Srivastava2021, Rajpurohit2020}. The spectrograph covers the spectral range of 4700 -- 6650$\AA$. \cite{Kumar2023} study was focused on the variability of H$\alpha$ and H$\beta$ emission lines over timescales ranging from $\sim$0.7 to 2.3 hours, with a cadence of approximately 3–10 minutes. In this study, a total of 126 M dwarfs were analysed in a spectral range of M0-M8.5. All the M dwarf sources used in this study show strong chromospheric activity, as indicated by H$\alpha$ equivalent widths less than -0.75 $\AA$, consistent with prominent H$\alpha$ emission \citep{Kumar2023}. For further details regarding the data reduction, the reader is referred to \cite{Kumar2023}.

\subsection{TESS photometry}
\subsubsection{Obtaining high cadence data}

The Transiting Exoplanet Survey Satellite (TESS) is a dedicated mission to monitor approximately 200,000 bright stars distributed across the sky. NASA launched TESS in April 2018. TESS observes in the spectral interval of 6000–10000 $\AA$ encompassing the blue optical to near-infrared regions. TESS satellites have provided unprecedented data with available 2-minute cadences for various stellar sources \citep{Ricker2015}. In this work, we have used TESS short cadence photometric data for the active M dwarfs in our sample. This high cadence photometric data enabled us to determine the $P_{\mathrm{rot}}$, FOR, flare energies, and $f_{\mathrm{s}}$ of the M dwarfs. Using the Python package lightkurve\footnote{\url{https://docs.lightkurve.org}}, we efficiently retrieved the light curves for these stars. The number of sectors per target ranges from 1 to 10. Complete sector lists along with TESS magnitudes and distances are provided in Appendix~\ref{tab:appendix_A1}. Out of 126 M dwarfs from our sample, 12 were excluded from this study, where 5 lacked TESS data, and 7 had no reliable light curves. For further details on the methodology and the determination of $P_{\mathrm{rot}}$ and $f_{\mathrm{s}}$, we refer to the study by \cite{Kumar2023}.    

\subsubsection{Quality checks and contamination}
For quality checks and contamination, we analyze 114 out of 126 M dwarfs from our sample, which were observed by TESS between 2018 and 2024 across multiple sectors and epochs. For consistency, we restrict our analysis to the light curves processed by the Science Processing Operations Centre (SPOC) \citep{Jenkins2016} and TESS-SPOC \citep{Caldwell2020} pipelines, adopting the Presearch Data Conditioning Simple Aperture Photometry (PDCSAP) flux. High-cadence observations with integration times of 20, 120, 200, and 600 seconds are available for subsets of the targets and are used in this study. For each target, we used all available TESS sector light curves for the analysis as provided in Appendix~\ref{tab:appendix_A1}. To quantify flux contamination from neighbouring sources, we employed TESSILATOR\footnote{\url{https://tessilator.readthedocs.io/en/latest/}}, which computes the total neighbour-to-target flux ratio $(\Sigma\eta)$, for more details, see \cite{Binks2024}. Targets with $\Sigma\eta>0.3$ were classified as contaminated and excluded from the flare analysis in this study. Using this criterion, out of 114 targets, 8 targets were identified as contaminated and removed from the final sample.\\

\begin{figure*}[hbt!]
	\centering
	\includegraphics[angle=0,height=0.35\textwidth,width=0.85\textwidth]{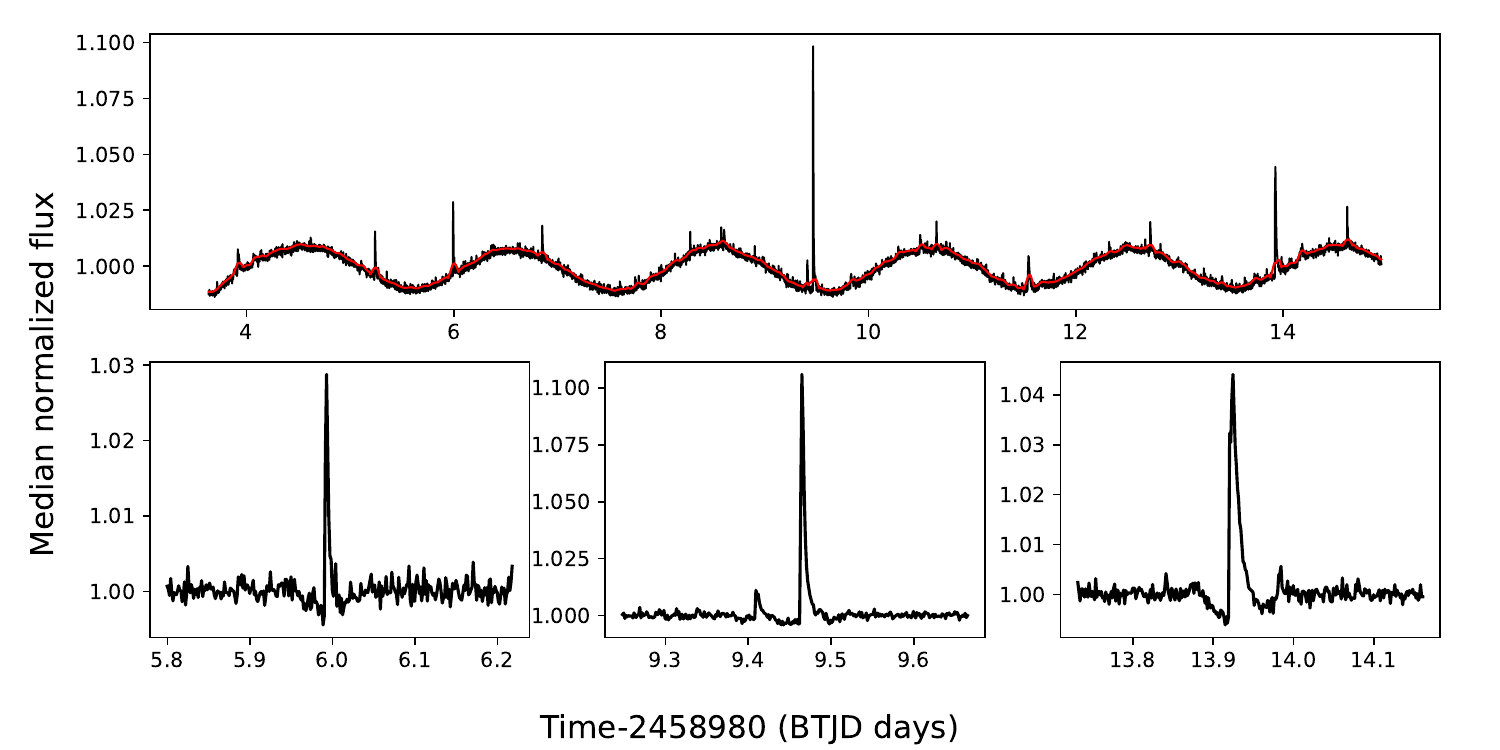}
	\caption{Top panel: TESS light curve of PM J16170+5516 (M2.0) from Sector-25, 2020 (SPOC) with $P_{\mathrm{rot}}$ 1.975 days. The black line shows the light curve with a cadence of 2 min, and the red line the Savitzky–Golay filtered and smoothed curve. Bottom panels: zoomed views of detrended flux highlighting multiple flare events of different magnitudes.}
	\label{fig2}
\end{figure*}

\section{Methods and analysis}

\subsection{Flare detection using photometric data from TESS}\label{flare detection}

Numerous algorithms for automated flare detection have been developed \citep{Walkowicz2011, Davenport2016, Gao2016}, and several methodologies have been applied to systematically identify and characterize these events in photometric light curves \citep{Feinstein2020b, Yang2023, Meng2023}. In this study, we employed the Altaipony\footnote{\url{https://altaipony.readthedocs.io}} Python package \citep{Ilin2021, Davenport2016} for the automated identification and analysis of flares within stellar light curves.\\

In the Altaipony module, initially, the de-trending of light curves was performed with a Savitzky-Golay filter \citep{Savitzky1964} to remove long-term trends. Then, the flare detection algorithm was employed using the Altaipony package. As outlined by \cite{Chang2015}, flares were identified based on three key parameters: N1, N2, and N3 within the Altaipony package. A candidate event was classified as a flare only if the following criteria are satisfied: 1) it exhibited a positive deviation from the median quiescent flux of the star, where the flux deviation at a given data point exceeds by N1$=$3 times the local scatter in the light curve, 2) the combined deviation and flux error exceed by N2$=$3 times the local scatter, 3) at least N3$=$3 consecutive data points had to satisfy the N1 and N2 thresholds.\\

Using this approach, 94 flaring objects were identified out of 106 M dwarfs observed with TESS. The complete lists of flaring and non-flaring stars are given in Appendix \ref{tab:appendix_A1} and \ref{tab:appendix_A2}. To quantify the robustness of our flare detections, using AltaiPony, we performed injection–recovery tests on the 20-s cadence TESS light curve of 2MASSWJ1012065-304926, an M6.0 star with a TESS magnitude of 14.5. Out of 58 injected synthetic flares, we recovered 51, corresponding to an overall completeness of $90\%$ with missed detections confined to the lowest-amplitude events. Our analysis with AltaiPony confirms the reliability of detecting moderate to strong flares, even in the lowest S/N case within our sample, with a false-positive rate $<5\%$. During the analysis, a visual inspection was also conducted by generating plots of flare candidates overlaid on the stellar background for each light curve, providing an additional layer of verification. Figure~\ref{fig2} illustrates the smoothed photometric light curve obtained during the de-trending process, where a Savitzky–Golay filter with a window length of 0.1 days and a second-order polynomial fit has been applied. One can see the quasi-sinusoidal brightness variations observed in PM J16170+5516 (M2.0), attributed to the rotational modulation caused by prominent star-spots coming into and out of view, as captured in TESS data \citep{Ricker2015}.

\subsection{Flare energy estimation}

M dwarfs are known for their frequent and energetic flaring activity. These transient events are associated with their fully convective envelopes, particularly in late-type stars with spectral types ranging from M3 to M9 \citep{Reiners2012a, Newton2017}. Flare energy is a critical parameter for quantifying the magnetic activity of stars. We calculated the energy of each flare following the methodology established by \cite{Shibayama2013} and \cite{Yang2017}, utilizing stellar luminosity, flare duration, and flare amplitude. Under the assumption that flaring M dwarfs act as blackbody radiators with an effective temperature of \( T_{\text{flare}} = 9000 \, \mathrm{K} \), as proposed by \cite{Kretzschmar2011}, the bolometric flare luminosity at a given time during the flaring event, $L_{\mathrm{flare, i}}$ is computed as:

$$
L_{\mathrm{flare}, i} = \sigma \, T_{\mathrm{flare}}^{4} \, A_{\mathrm{flare}, i}
$$

where \(\sigma\) denotes the Stefan–Boltzmann constant, and $A_{\mathrm{flare}, i}$ represents the flare area, and is determined by the relationship:

$$
A_{\mathrm{flare}, i} = C_{\mathrm{flare}, i} \, \pi R^{2} 
\frac{\displaystyle \int R_{\lambda} B_{\lambda}\!\left(T_{\mathrm{eff}}\right) \, d\lambda}
{\displaystyle \int R_{\lambda} B_{\lambda}\!\left(T_{\mathrm{flare}}\right) \, d\lambda},
$$

where $C_{\mathrm{flare}, i}$ is defined as:

$$
C_{\mathrm{flare}, i} = 
\frac{F_{i} - F_{o}}{F_{o}}
$$

In this formulation, $R_{\lambda}$ is the TESS response function\footnote{\url{https://heasarc.gsfc.nasa.gov/docs/tess/the-tess-space-telescope.html}}, \( B_\lambda(T) \) is the Planck function at temperature \( T \), \( R \) is the stellar radius, \( F_{\text{i}} \) is the measured flux at a given time during the flare event, and \( F_{\text{o}} \) is the quiescent flux (at the same time) of the star. Stellar radii were taken directly from the TESS Input Catalog (TIC) and Candidate Target List (CTL) when available \citep{Stassun2018} and are tabulated in appendix \ref{tab:appendix_A1}. The total bolometric energy of a flare is then obtained by integrating $L_{\mathrm{flare}, i}$ over the whole duration of the flare, whereas the flare duration is determined using the Altaipony algorithm. The uncertainties in the flare energy estimations are significant, with \cite{Shibayama2013} reporting an approximate error margin of \(\pm 60\%\). The estimated stellar radius and the average flare energy $\langle E_{\mathrm{flare}} \rangle
= \frac{1}{N}\sum_{i=1}^{N} E_i$, and $L_{\mathrm{flare}} / L_{\mathrm{bol}}$ are tabulated in appendix \ref{tab:appendix_B}.

\subsection{Flare occurrence rate (FOR), flare duration, and flare amplitude}
 
In this study, we also examined the relationship between the $P_{\mathrm{rot}}$ and the FOR within our sample of M dwarfs. For sources having flare episodes, the FOR is calculated as the ratio of the total duration of all flaring events to the total observing duration defined by \cite{Walkowicz2011}.

\[
\mathrm{FOR} = 
\frac{\displaystyle \sum t_{\mathrm{flare}}}
{\displaystyle \sum t_{\mathrm{star}}} \times 100
\]

Where \(\sum t_{\text{flare}}\) is the summation of the flare duration of each star and \(\sum t_{\text{star}}\) is the observation period of each star. The flare duration is determined using the Altaipony algorithm, and the observation period was estimated by multiplying the exposure time of each data point by the total number of data points searched for flare epochs in the de-trended light curve.\\

Flare amplitude is also a key parameter defining a flare event's characteristics. To understand the flare amplitude statistics across the spectral range in our sources, the flare amplitude has been determined for each flare using the Altaipony algorithm, which measures the difference in flux at the peak of the flare relative to the quiescent stellar flux. FOR and median flare amplitude for each source are tabulated in appendix \ref{tab:appendix_B}.\\

\subsection{Flare frequency distribution}\label{Analysis:FFD}
As flares span a wide energy range, cumulative Flare frequency distributions (FFDs) are commonly used to describe the occurrence of flares as a function of energy \citep{Hawley2014, Davenport2016, Ilin2021, Jackman2021}. The cumulative flare frequency $\nu$ $\sim$ $ (>E)$ for a given flare energy $E$ is determined by the total number of flares with flare energy $\geq E$, divided by the total observation duration of the target. FFDs follow power-law relations and are written as

\[
dN(E) = k \, E^{-\alpha} \, dE
\]

where $N$ is the number of flares that occur in the total observation duration, $E$ is the flare energy, $k$ is a proportionality constant, and $\alpha$ is the power law index. Thus, the flare frequency ($\nu$) can be estimated by integrating the above equation,

\[
\log \nu = 
\log\!\!\left(\frac{k}{1-\alpha}\right) 
+ (1-\alpha)\,\log E
\]

The parameter $\alpha$ determines the relative occurrence of high-energy flares compared to lower-energy ones and has been found to vary across different spectral types. Previous studies have shown that $\alpha$ is negative, indicating that high-energy flares are less frequent than their low-energy counterparts \citep{Shibayama2013}.\\

The low-energy regime of FFDs is often incomplete due to the inability to recover all low-energy flares \citep{Davenport2016}. This incompleteness can be noticed as a deviation from the expected power-law behavior, appearing as a flattening in log-log space at lower energies. The detection method becomes less sensitive at these low energies, resulting in an underestimation of flare detections. To mitigate this effect, many studies have restricted their analysis to flare energies above a certain threshold, well beyond the turnover, ensuring a more reliable power-law fit \citep{Hawley2014}. Following the same approach, we also restrict the data for power-law fitting to the energy of 10$^{32}$ ergs (see Table \ref{tab:ffd} for more details). Section \ref{Flare_results} examines the FFDs of M dwarfs by categorizing them based on spectral type using this power-law approach.\\

\section{Results and Discussion}\label{Result}
\subsection{H$\alpha$ variability}

Various observable phenomena occurring in the outer stellar atmosphere, such as strong stellar winds, flares, coronal mass ejections, and star-spots, are generally used to characterize the magnetic activity in stars. These processes give rise to distinct chromospheric emission lines, among which the H$\alpha$ emission line is widely utilized for activity-related studies, particularly in M dwarfs. Investigating the short-term variability of H$\alpha$ emission line properties provides valuable insight into these magnetic processes. The statistical properties of H$\alpha$ equivalent width and H$\alpha$ activity strength ($L_{\mathrm{H\alpha}} / L_{\mathrm{bol}}$) were previously analyzed in \cite{Kumar2023} by spectroscopic monitoring for a sample of M dwarfs. For a detailed discussion on the estimation methods and statistical analysis of these parameters, we refer the reader to \cite{Kumar2023}. In this study, we have used these properties to find a plausible relationship between chromospheric activities and the flare phenomenon. 

\begin{figure}[hbt!]
	\includegraphics[angle=0,width=0.48\textwidth]{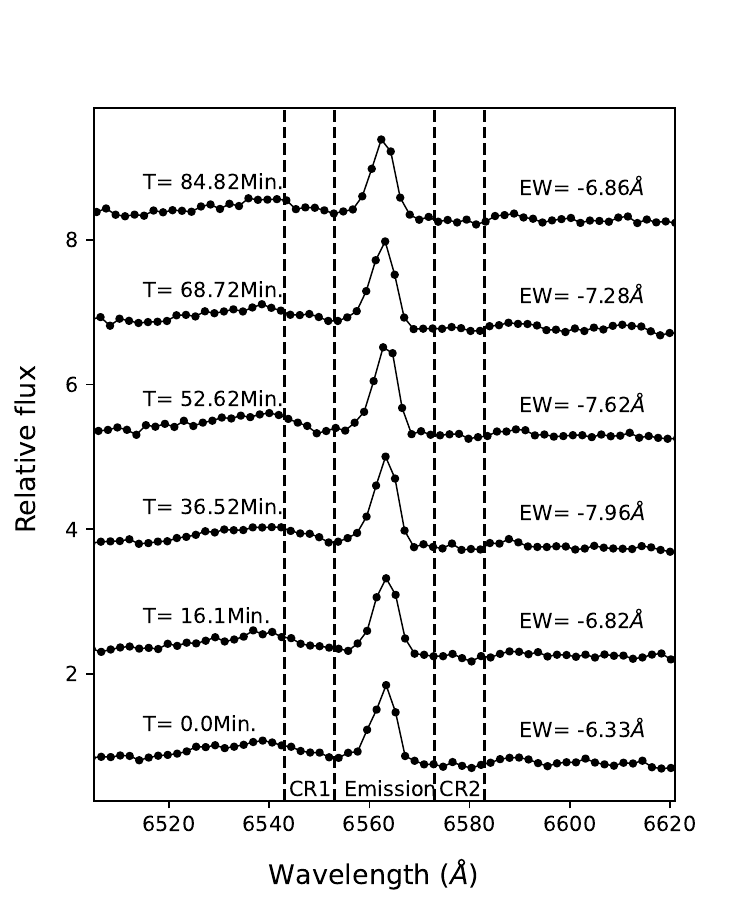}
	\caption{Temporal variation in the H$\alpha$ line profiles for PM J12142+0037 (M4.0) since the beginning of the first exposure. Dashed lines indicate the pseudo-continuum regions. Equivalent widths are marked to the left of each emission line, and elapsed times since the initial exposure are shown on the right.}
	\label{fig3}
\end{figure}

Figure~\ref {fig3}, as an example, showcases various temporal spectra from different observations of flaring M dwarf PM J12142+0037 of spectral type M4. One can see the temporal changes in the H$\alpha$ emission since the beginning of the first exposure. This enhanced emission of H$\alpha$ is generated through collisional excitation within the relatively dense chromosphere and is the most prominent and commonly utilized indicator of stellar magnetic activity. Such temporal variation in the H$\alpha$ emission line could be linked to the emergence of active regions on the stellar surface.

\begin{figure}[hbt!]
	\centering
	\includegraphics[angle=0,width=0.48\textwidth]{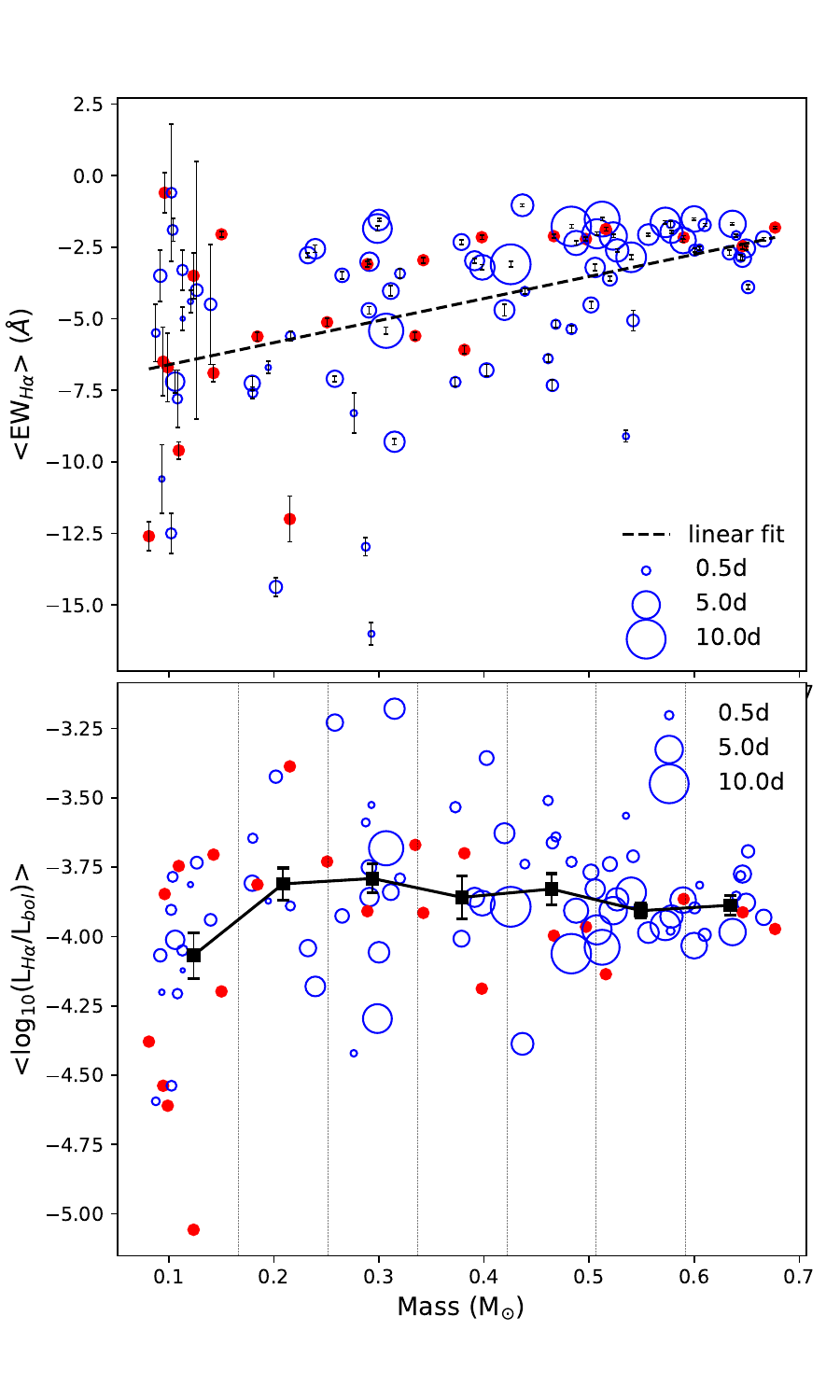}
    \vspace{-1.0cm}
	\caption{Variability and activity indicators as a function of stellar mass. Top panel: distribution of $\Delta \mathrm{EW} = \max(\mathrm{EW}) - \min(\mathrm{EW})$ for H$\alpha$ emission. The black dashed line shows a linear fit, $ \Delta \mathrm{EW}_{\mathrm{H}\alpha} = a(M_\ast) + b$, with    $a = 7.71 \pm 1.55$, \quad $b = -7.38 \pm 0.64$. Bottom panel: median values of $\log_{10}(L_{\mathrm{H}\alpha}/L_{\mathrm{bol}})$ for H$\alpha$ emission as a function of stellar mass. Black squares mark the medians in seven equally spaced bins, with scatter quantified using the median absolute deviation (MAD) \ref{Results:variability}. Blue circles are scaled by $P_{\mathrm{rot}}$, with larger circles corresponding to longer $P_{\mathrm{rot}}$. The red-filled circles indicate stars with no measured rotation period. Vertical gray lines show the bin sizes.}
	\label{fig4}
\end{figure}

\begin{figure*}[hbt!]
	\begin{center}
	\includegraphics[width=0.95\textwidth]{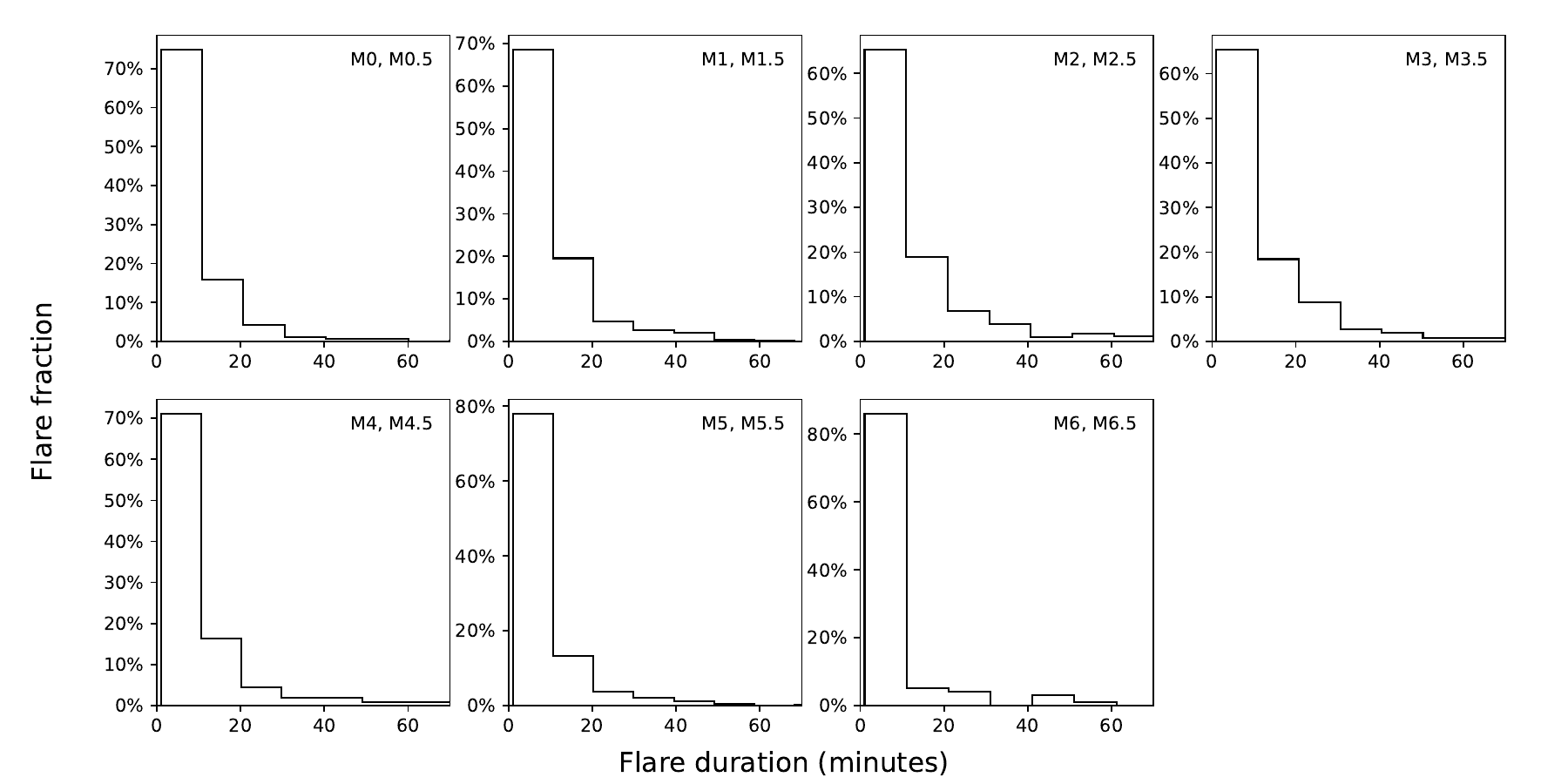}
    \end{center}
    \vspace{-0.5 cm}
	\caption{Distribution of flare durations for M dwarfs (in minutes) shown across spectral subtypes M0–M6.5}
	\label{fig5}
\end{figure*}

\begin{figure*}[h!]
\begin{center}
\includegraphics[width=0.95\linewidth]{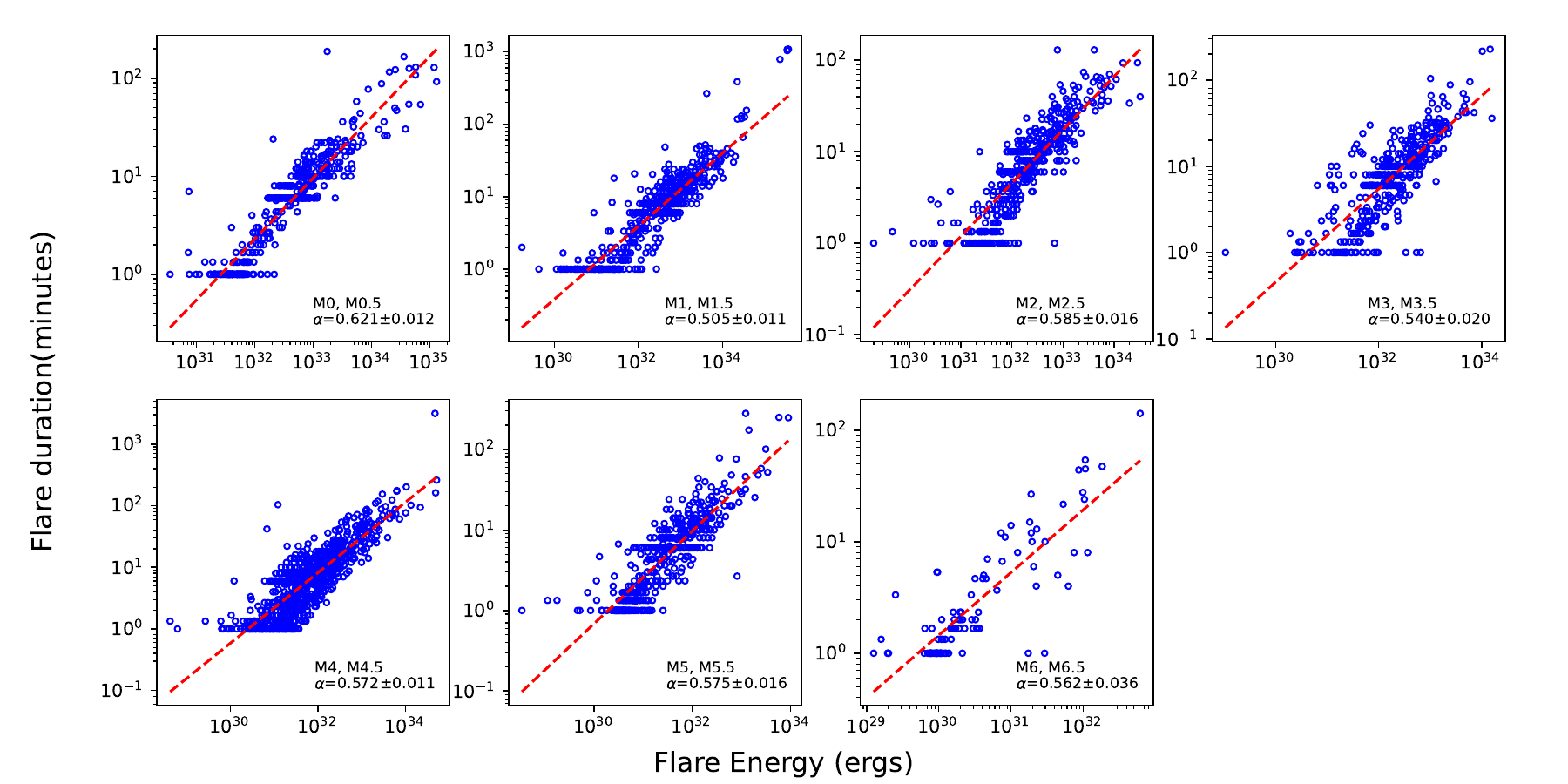}
\end{center}
\vspace{-0.5 cm}
\caption{Flare energy–duration relation for M dwarfs. Red dashed line shows the power-law fit (see Section~\ref{Analysis:FFD}).}
\label{fig6}
\end{figure*}

\subsection{Chromospheric activity as a function of mass}\label{Results:variability}

Studies by \cite{Silva2011} and \cite{Robertson2013} suggested that magnetic activity is higher in massive M dwarfs compared to their lower-mass counterparts. Notably, \cite{Robertson2013} proposed that compared to either effective temperature (T$_{\mathrm{eff}}$) or stellar color, stellar mass serves as a better predictor of mean H$\alpha$ activity. The relationship between chromospheric activity and mass in M dwarfs is more complex than in massive stars. However, some correlation between chromospheric activity and mass may still exist within the M dwarf population.

To explore the possibility of any correlation between chromospheric activity strength and stellar mass among flaring M dwarfs in our sample, we adopt stellar masses provided by \cite{Stassun2018}. They are derived using the $M_{K_\mathrm{s}}$–mass relation from \citet{Mann2015} and \citet{Benedict2016}. In the top panel of Figure~\ref{fig4}, we display the distribution of $\Delta \mathrm{EW} = \max(\mathrm{EW}) - \min(\mathrm{EW})$ for H$\alpha$ emission as a function of seven equally spaced mass bins. Our analysis shows that for the lower mass regime \(M_\odot\) $<$ 0.2, H$\alpha$ shows larger scatter but decreases steadily at higher masses \(M_\odot\) $>$ ~0.35. The bottom panel of figure~\ref{fig4} shows the distribution of median values of $L_{\rm H\alpha}/L_{\rm bol}$ as a function of stellar mass, where the scatter in each bin is quantified using the Median Absolute Deviation (MAD), scaled by 1.48 for consistency \citep{Mazeh2015}. The uncertainty in the median for each bin is estimated as $\frac{\text{MAD}}{\sqrt{N}}$, where N represents the number of data points within the bin. We find that $L_{\rm H\alpha}/L_{\rm bol}$ peaks nearly at $\sim$ 0.2-0.3\(M_\odot\) and then becomes nearly flat beyond $\approx 0.3~M_{\odot}$. We observe that $\Delta{\rm EW}$ decreases by approximately 92$\%$, spanning a factor of about 6.3 times above and below the linear fit. Such variation indicates that despite their long activity lifetimes, the coolest M dwarfs exhibit reduced chromospheric heating efficiency. This indicates chromospheric activity across the mass range of $\sim$0.2-0.3\(M_\odot\) may be attributed to the transition to fully convective interiors \cite{Reiners2012a, West2015, Newton2017}. Such changes induce fundamental changes in the processes responsible for generating strong magnetic fields in M dwarfs. However, the limited sample size prevents us from drawing firm conclusions. Nevertheless, we infer that the variability in the chromospheric H$\alpha$ emission exhibits a positive correlation with stellar mass, consistent with the findings of \cite{Robertson2013, Lin2019} and \cite{Gunther2020}.

\subsection{Flare energies, flare duration, and flare frequency distribution}\label{Flare_results}

As discussed in Section \ref{flare detection}, 95 M dwarfs out of 106 sources observed by TESS exhibit flare episodes. Flare duration, amplitude, and energies have been estimated for these sources. In figure~\ref{fig5}, we show the distribution of flare-event duration, suggesting that flares that last for less than 10 minutes are $\sim$65-86$\%$, while there are very few ($\sim$2-4\%) flaring events lasting more than 60 minutes. We find that short-duration flares ($<$ 10 minutes) are significantly more common in M dwarfs than long-duration flares ($>$ 10 minutes). We see this trend across all spectral subtypes of M dwarfs, with later-type M dwarfs (M4-M6.5) exhibiting the highest relative frequency of short-duration flares. The common nature of these rapid flares suggests that magnetic reconnection events in M dwarfs often result in impulsive energy releases, likely influenced by their fully convective nature and strong magnetic activity. Additionally, this high incidence of short-duration flares is closely linked to the rapid rotation rates of M dwarfs as it enhances the stellar activity by increasing dynamo efficiency, leading to more frequent and shorter-duration flares \citep{Newton2017, West2015, Lin2019, Gunther2020}. Observational studies from Kepler, TESS, and K2 show that the majority of detected flares in M dwarfs are of short duration, with only a small fraction persisting beyond $\sim$20 minutes \citep{Davenport2014, Hilton2010, Raetz2020}. The decline in longer-duration flares across all subtypes implies that large-scale energy storage and release events are less frequent. Our results are consistent with the fact that magnetic reconnection events in M dwarfs are influenced by stellar structure, with fully convective stars (M4 or later) showing more frequent but shorter flares compared to partially convective stars (M0–M3). The details of flaring sources, number of flares, and flares percentage with t$_{\rm flare}<$ 10 minutes are tabulated in Table \ref{tab:ffd}.

\begin{figure*}[!thp]
\begin{center}
\includegraphics[width=0.95\linewidth]{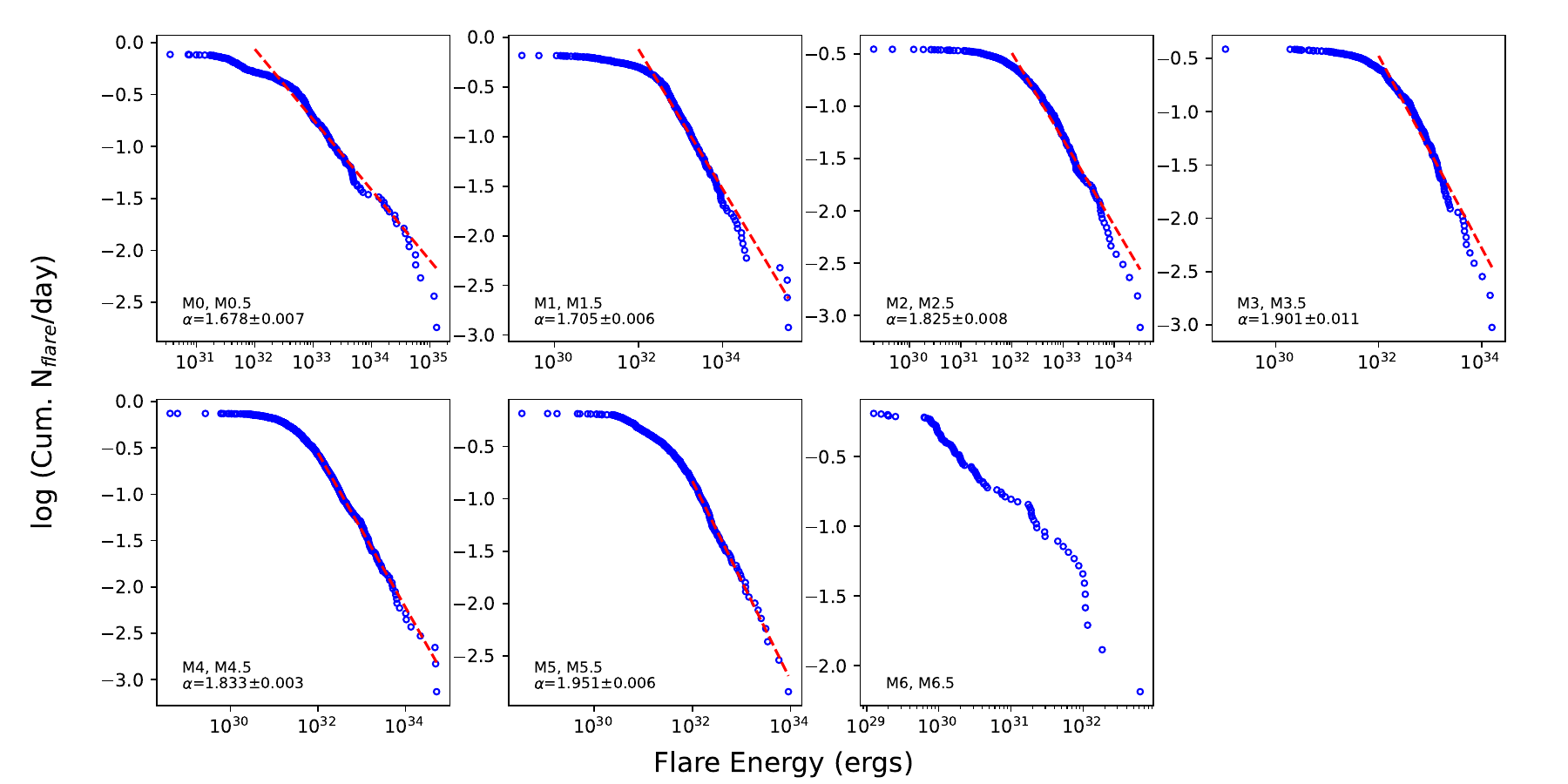}
\end{center}
\vspace{-0.5 cm}
\caption{Cumulative flare frequency distribution for M dwarfs plotted as a function of flare energy across spectral types. The red dashed line shows the power-law fit (see Section~\ref{Analysis:FFD}). For the M6–M6.5 sample, no fit is shown due to the limited number of flares.}
\label{fig7}
\end{figure*}

Using TESS, Kepler, and K2 data, various previous studies \citep{Shibayama2013, Davenport2016, Lin2019, Zhang2020} founds that the flare energy ranges between $\sim$ $1 \times 10^{33}$ and $1 \times 10^{36}$ ergs with an upper limit of $1 \times 10^{38}$. Based on TESS data, the study by \cite{Doyle2019} found that flare energies in M dwarfs range from $6 \times 10^{29}$ to $2.4 \times 10^{35}$ ergs, whereas the study by \cite{Gunther2020} flare energy ranges from $1 \times 10^{31}$ to $10^{38}$. As suggested by \cite{Yang2023}, we also found that there is a clear boundary at about $10^{35}$ erg, and the flare energies of our M dwarf stars are less than $10^{35}$ erg. The estimated average energy that is average energy per flare $\langle E_{\mathrm{flare}} \rangle$ for each source is tabulated in appendix \ref{tab:appendix_B}. The relationship between flare energy and flare duration for different M sub-spectral types is plotted in Figure~\ref{fig6}. The dashed lines represent the power-law fit. The value of $\alpha$ lies between approximately 0.505 and 0.621, indicating that higher-energy flares generally have longer durations. We attribute the variation in $\alpha$ across spectral types to the different magnetic field structures and their different convective nature. Our findings are consistent with studies that have examined the rotation-activity relations and flare characteristics of M dwarfs \citep{Lin2019, Doyle2019, Yang2023}.

As we have all the flare energies in our analysis for the stars, the FFDs for the flare energies can better describe the behavior of flare activity in stars. We computed the FFDs for M dwarfs, categorizing them into spectral subgroups, as explained in section \ref{Analysis:FFD}. In figure~\ref{fig7}, we presented the cumulative flare frequency distribution of flaring M dwarfs of different M sub-spectral types overlaid with a power law fit (dashed line). For the M6–M6.5 sample (last panel), no power-law fit is shown due to the limited number of flares. We restrict our power-law fitting across all spectral type bins to flares with energies above $10^{32}$ following the approach of \citet{Hawley2014}. The coefficient $\alpha$, as described in section \ref{Analysis:FFD}, provides insight into the star's flare activity. We find that $\alpha$ is $\approx 1.68$ for M0-M0.5, whereas for M5-M5.5, $\alpha$ is  $\approx 1.95$, indicating that the distribution of flare energies changes systematically from earlier to later spectral types. It is evident from the figure that low-energy flares dominate the overall distribution, while high-energy flares become progressively rarer. Also, the slightly steeper slopes for M dwarfs M4 to M5 could be due to a larger contribution from low-energy flares. The dependence $\alpha$ across spectral subtypes suggests that the underlying flare generation mechanism is broadly coherent \citep{Lin2019}.

\subsection{Relationship of FOR with $\teff$, $P_{\mathrm{rot}}$ and Flare amplitude}
 
Stellar flares occur spontaneously and are difficult to predict. In M dwarfs, the detection of flares is very challenging because of their sporadic nature and the observational limitations of ground-based telescopes. The FOR values vary across spectral types, reaching their peak in stars with strong magnetic dynamos and decreasing in those with weaker magnetic fields. Compared to G and K dwarfs, cooler M dwarfs sustain more active dynamos, leading to a higher FOR during their early, fast-rotating stages. 

\begin{figure}
	\centering
	\includegraphics[angle=0,width=0.48\textwidth]{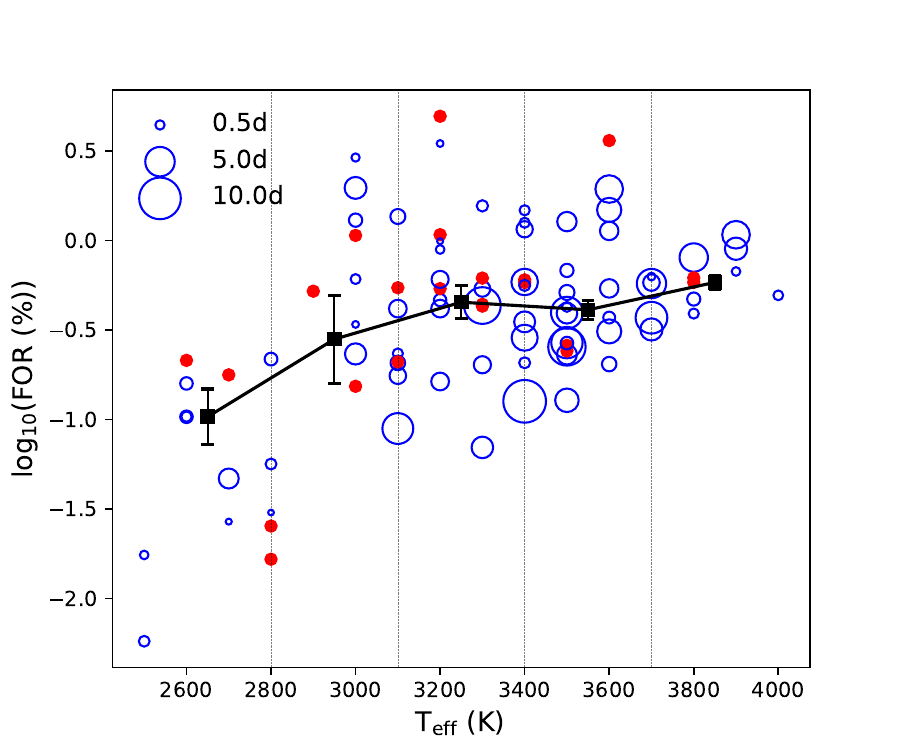}
	\caption{FOR versus $\teff$ for M dwarfs. Black squares are the median values in five equally spaced bins. Symbols have the same meaning as in Fig.~\ref{fig4}.}
	\label{fig8}
\end{figure}

Figure~\ref{fig8} presents the relationship between $\teff$ and FOR. Our analysis shows that FOR increases with $\teff$ up to $\sim$ 3200 K, above which it increases slowly. This observed rise in FOR in M dwarfs with $\teff$ is consistent with the fact that hotter M dwarfs (M0-M4) have stronger magnetic and chromospheric activity than cooler M dwarfs (M4 or later). Also, as suggested by \cite{Meng2023}, lower-intensity flares are more easily detected against the lower continuum of cooler M dwarfs. Our results may not be statistically significant as the number of flaring stars in the spectral range M5-M9 is limited, and we need more M dwarfs in that spectral range. Nevertheless, our findings are aligned with previous studies by \cite{Lin2019, Stelzer2022} and \cite{Yang2023}.

\begin{figure}
	\includegraphics[width=0.46\textwidth]{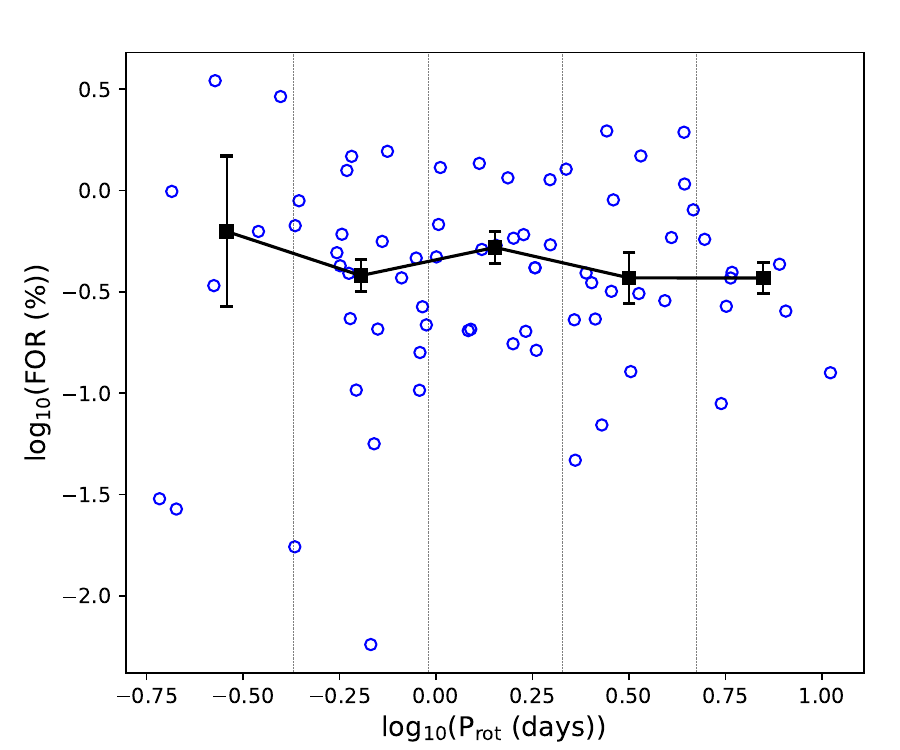}
	\caption{FOR as a function of $P_{\mathrm{rot}}$. Blue circles represent individual stars. Black square marks the median values in five equally spaced bins on a log scale. Error bars on the medians are discussed in Section \ref{Results:variability}.}
	\label{fig9}
\end{figure}

\begin{figure}
	\centering
	\includegraphics[width=0.48\textwidth]{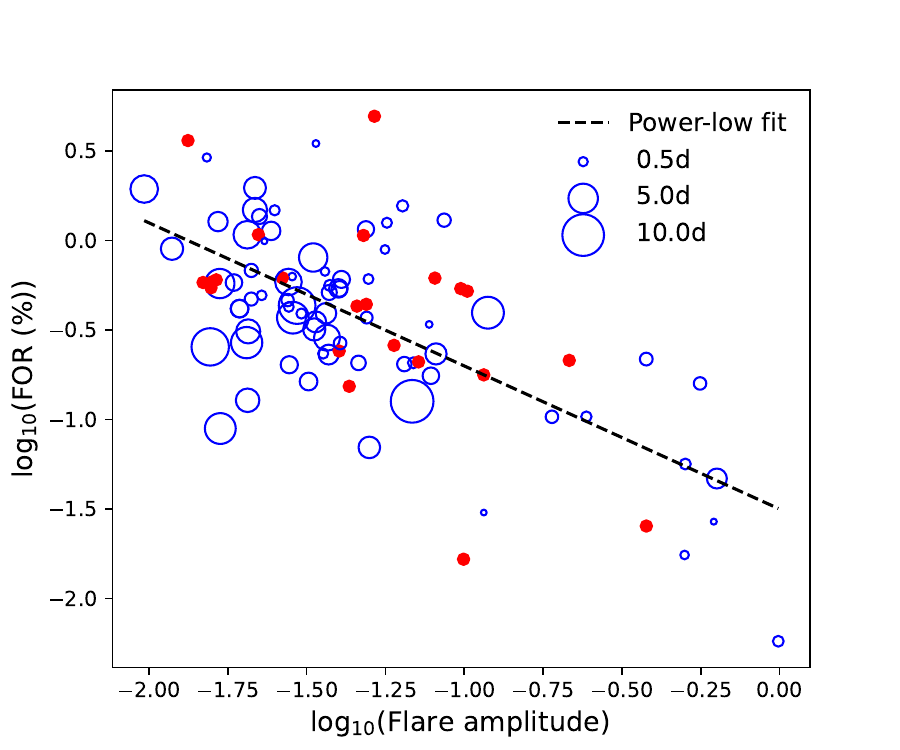}
	\caption{Flare amplitude versus FOR. Dashed line shows the best-fit power law, $\mathrm{FOR} = k\,(\text{Flare Amplitude})^{\alpha}$, where $k = 0.032 \pm 0.01$ and $\alpha = -0.8 \pm 0.09$. Symbols have the same meaning as in Fig.~\ref{fig4}.}
	\label{fig10}
\end{figure}

\begin{figure}
    \centering
    \includegraphics[width=0.48\textwidth]{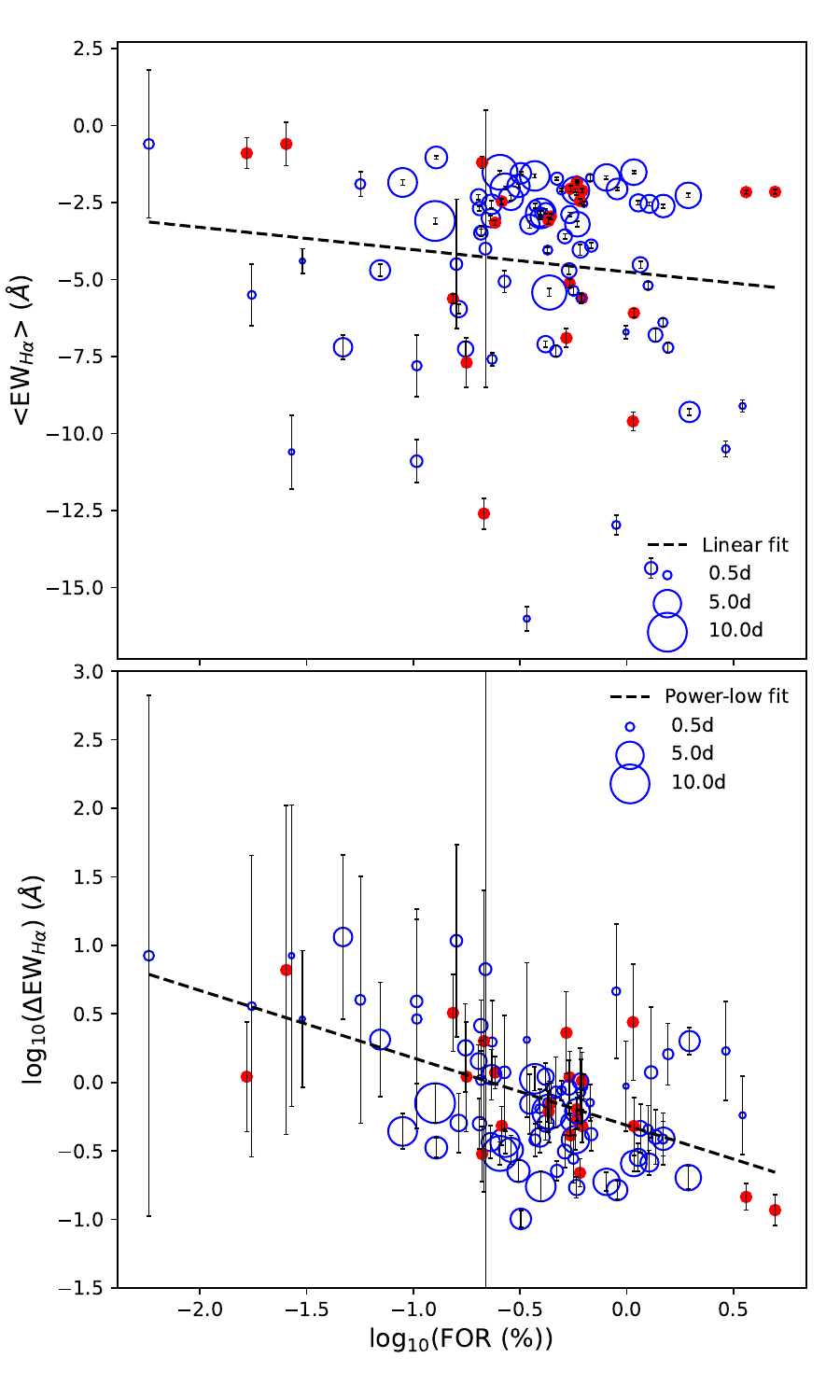} 
    \caption{FOR versus H$\alpha$ activity indicator. Upper panel: linear fit as $\langle \mathrm{EW}_{\mathrm{H}\alpha} \rangle = a \,[\log_{10}(\mathrm{FOR})] + b$, where $a = -0.72 \pm 0.63$ and $b = -4.75 \pm 0.42$; Lower panel: power-law fit as $\Delta\mathrm{EW}_{\mathrm{H}\alpha} = k\,(\mathrm{FOR})^{\alpha}$, where $k = 0.485 \pm 0.058$ and $\alpha = -0.492 \pm 0.076$. Symbols have the same meaning as in Fig.~\ref{fig4}.}
    \label{fig11}
\end{figure}

Stellar rotation plays a critical role in determining the flaring activity of M dwarfs. It is one of the essential parameters that affect flare events. In general, faster-rotating M dwarfs exhibit significantly higher FOR than slower rotators \citep{Wright2011, Notsu2013, Candelaresi2014, Yang2023}. Since all M dwarfs in this study have measurable star-spots indicative of strong magnetic activity, they are observed to produce flares from TESS light curves. For this study, we have used the $P_{\mathrm{rot}}$ derived by \cite{Kumar2023}. The sample comprises fast-rotating M dwarfs with $P_{\mathrm{rot}}$ $<$ 10 days. The $P_{\mathrm{rot}}$ are grouped into bins, with $P_{\mathrm{rot}}$ of each bin used for the analysis. The uncertainty in the median is calculated as explained in section \ref{Results:variability}. It is important to note that individual TESS sectors span approximately 27 days, which limits the sensitivity of period detection to roughly 10 days or shorter.

In figure~\ref{fig9} and figure~\ref{fig10}, we explore the relationship between the $P_{\mathrm{rot}}$ with FOR and flare amplitude. The $P_{\mathrm{rot}}$ of M dwarfs in our sample range from 0.12 to 10.5 days. From figure~\ref{fig9}, we find that FOR remains constant across the $P_{\mathrm{rot}}$ $<$ 10 days. Though the individual stars show larger scatter, the binned averages do not show a significant dependence of FOR on $P_{\mathrm{rot}}$. We infer that such behavior could be due to their efficient dynamos that can sustain frequent flaring activity in the fast-rotating ($P_{\mathrm{rot}}$ < 10 days) M dwarfs. While our findings align with previous studies, including \cite{Lin2019} and \cite{Yang2023}, we note that our sample is limited to fast-rotating M dwarfs $P_{\mathrm{rot}}$ $<$ 10 days. Further investigation is needed to explore whether such behavior persists in M dwarfs with FOR in M dwarfs with $P_{\mathrm{rot}}$ $>$ 10 days.
 
In Figure~\ref{fig10}, we show the median of peak flare amplitudes ($A_{\mathrm{Med}}$) of all flare events for each source versus FOR. The flare median amplitudes are usually found to be smaller and in the range of 0.01-1.0, and flare amplitudes are independent of stellar rotations. We also find a negative correlation between FOR and flare amplitude. This could be because M dwarfs with higher FOR tend to exhibit lower flare amplitudes, indicating that frequent flares in these M dwarfs are generally less energetic \citep{Raetz2020}. Also, more active stars release their magnetic energy in smaller, frequent bursts, whereas less active stars accumulate magnetic energy over longer periods, leading to rarer but more powerful flares \citep{Stelzer2016, Yang2023, Gunther2020}.

Several theoretical models and observational studies have established a strong correlation between stellar flares and chromospheric activity. The H$\alpha$ emission line, which is highly sensitive to magnetic field strength and $P_{\mathrm{rot}}$, is a key diagnostic of chromospheric activity and is closely linked to flare mechanisms \citep{Chang2017, Zhang2020}. We examine the connection between FOR and chromospheric activity indicators to explore further this relationship, specifically the H$\alpha$ equivalent width (EW) and its variability, $\Delta(\mathrm{H}\alpha\ \mathrm{EW})$, as illustrated in Figure~\ref{fig11}. Our analysis reveals a slight negative correlation between H$\alpha$ EW and $\Delta(\mathrm{H}\alpha\ \mathrm{EW})$
 as a function of FOR, with M dwarfs exhibiting larger H$\alpha$ EW and $\Delta$(H$\alpha$ EW) tending to have higher FOR. Strong and variable H$\alpha$ emission suggests an actively heated chromosphere, driven by underlying magnetic fields, which contributes to frequent and energetic flaring activity \citep{Doyle2019, Lin2019}. Given that the M dwarfs in this study predominantly exhibit $P_{\mathrm{rot}}$ $<$ 10 days, their high FOR can be attributed to rapid rotation, which amplifies their magnetic dynamo efficiency and sustains heightened magnetic activity \citep{Newton2017}.

\subsection{Flare activity: quantification}

According to the stellar dynamo theory, the interaction between stellar rotation and convection drives the generation of magnetic fields. The stellar chromospheric activity is closely linked to the rotation period. \cite{Noyes1984} studied the effect of these two factors on magnetic dynamo efficiency and combined them into a single parameter known as the Rossby number.  

\begin{figure}[hbt!]
	\includegraphics[angle=0,width=0.45\textwidth]{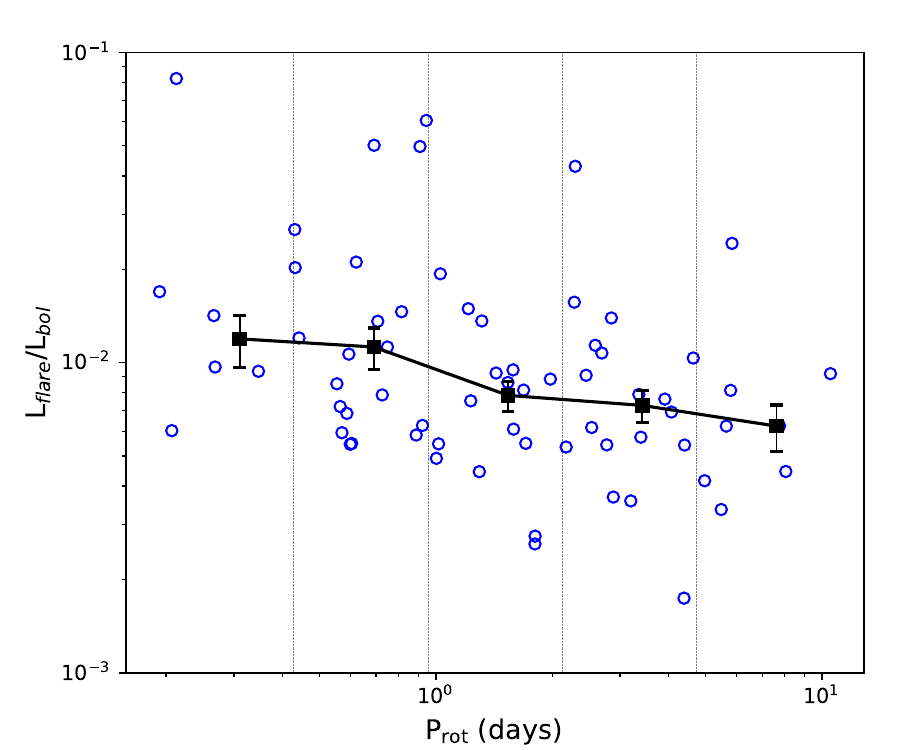}
	\caption{Distribution of $L_{\mathrm{flare}} / L_{\mathrm{bol}}$ with $P_{\mathrm{rot}}$ for M dwarfs. Blue circles represent individual stars. Black square symbols represent the median values in five equally spaced bins (log scale). The error bars in the median values are discussed in section \ref{Results:variability}.}
	\label{fig12}
\end{figure}

In figure~\ref{fig12}, we presented the relationship between the flare activity ($L_{\rm flare}/L_{\rm bol}$) and the $P_{\mathrm{rot}}$. As expected for M dwarfs, we see a similar trend where there is a decline in $L_{\rm flare}/L_{\rm bol}$ as $P_{\mathrm{rot}}$ increases. From our analysis, unlike traditional saturation observed in other activity indicators as shown by \cite{Yang2017}, we do not see such saturation because we do not have M dwarfs with $P_{\mathrm{rot}}$ $>$ 10 days. 

\begin{figure}[hbt!]
	\centering
	\includegraphics[angle=0,width=0.48\textwidth]{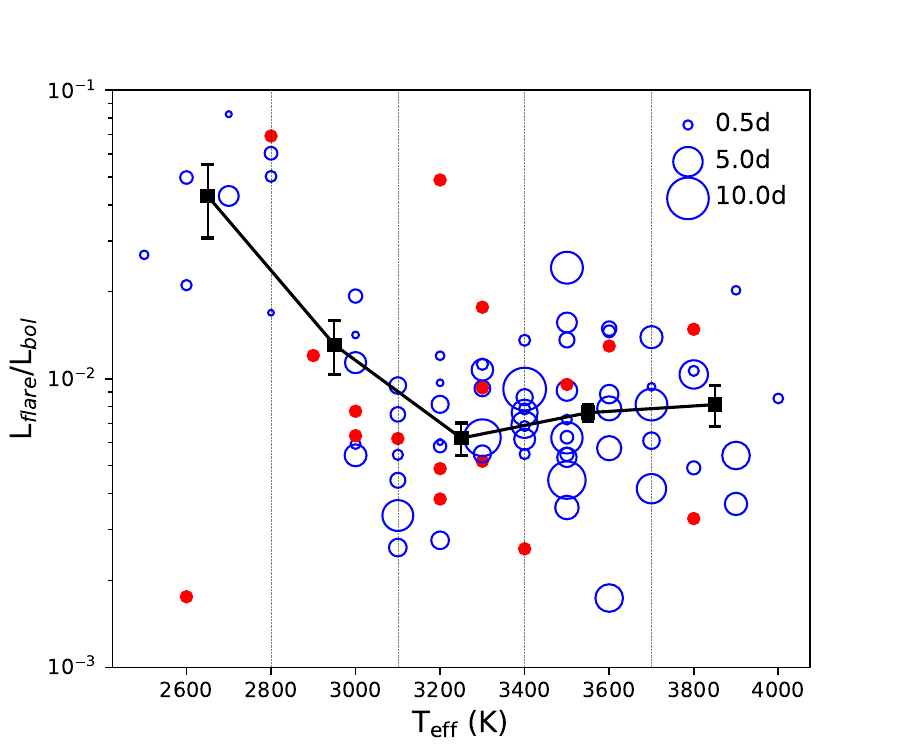}
	\caption{Distribution of $\teff$ with $L_{\mathrm{flare}} / L_{\mathrm{bol}}$. Black square symbols represent the median values in five equally spaced bins. Symbols have the same meaning as in Fig.~\ref{fig4}.}
	\label{fig13}
\end{figure}

The relationship between $L_{\rm flare}/L_{\rm bol}$ and $\teff$ is depicted in figure~\ref{fig13}. We observe that the flare activity for early-type M dwarfs (M0-M4) with $\teff$ $>$ 3200 K remains relatively stable around $\sim$$5 \times 10^{-3}$. In contrast, a rapid increase of nearly an order of magnitude in the average flare activity is observed for M dwarfs with $\teff$ $<$ 3200 K, indicating the disappearance of low-activity flaring stars at cooler $\teff$. At the same $\teff$, stars with shorter $P_{\mathrm{rot}}$ generally exhibit stronger flare activities. However, we would like to mention that the overall number of M dwarfs with $\teff$ $<$ 3200 K is much less than in other warmer $\teff$ bins.

\begin{figure}[hbt!]
	\centering
	\includegraphics[width=0.45\textwidth]{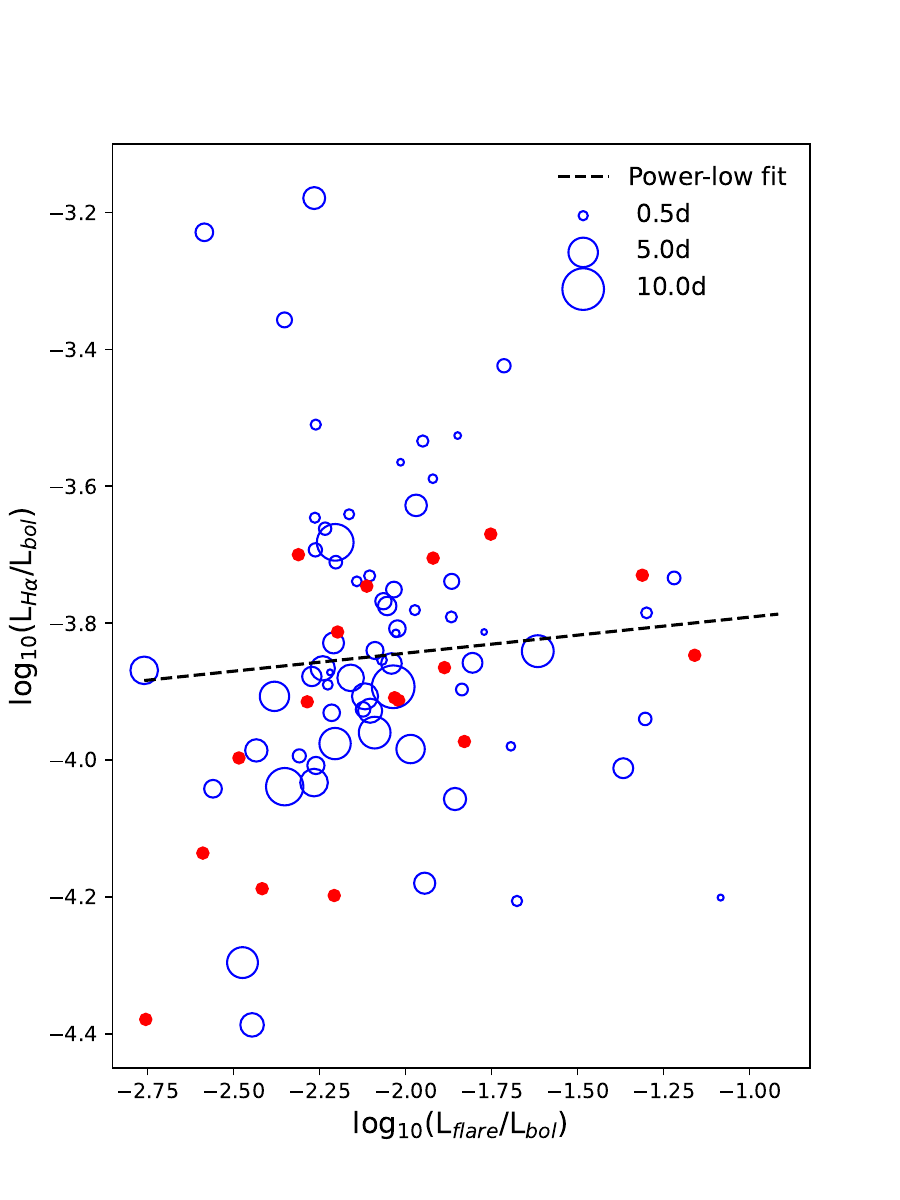}
	\caption{Flare activity ($L_{\rm flare}/L_{\rm bol}$) versus H$\alpha$ luminosity ($L_{\rm H\alpha}/L_{\rm bol}$). Black-dashed line shows power-law fitting as $L_{\text{H}\alpha} / L_{\text{bol}} = k\left( L_{\text{flare}} / L_{\text{bol}} \right)^\alpha$ where k=(1.83$\pm$0.63)$\times$10$^{-4}$ and $\alpha$=0.053$\pm$0.072. Symbols have the same meaning as in Fig.~\ref{fig4}.}
	\label{fig14}
\end{figure}

In active stars, the H$\alpha$ emission line is highly affected by the magnetic fields and $P_{\mathrm{rot}}$, which play a crucial role in the flare mechanism. As the H$\alpha$ luminosity ($L_{\rm H\alpha}/L_{\rm bol}$) and flare activity ($L_{\rm flare}/L_{\rm bol}$) also serve as a proxy for stellar activity, examining the relationship between them is very useful. Thus, we have also attempted to examine the relationship between the H$\alpha$ luminosity ($L_{\rm H\alpha}/L_{\rm bol}$) and flare activity ($L_{\rm flare}/L_{\rm bol}$). Figure~\ref{fig14} illustrates that $L_{\rm H\alpha}/L_{\rm bol}$ generally follows a power-law dependence on flare activity. The long-period stars ($P_{\mathrm{rot}}$ $>$ 5 days) have low H$\alpha$ luminosity on average. There are also short-period stars with low H$\alpha$ luminosity. The figure also suggests that flare activity may have a stronger dependence on the rotation period than chromospheric activity, as observed by \cite{Yang2017}. This relationship is important as it reveals the connection between chromospheric activity and energy release from the photosphere.

\subsection{Star-spot filling factor ($f_{\mathrm{s}}$)}

It is well known that the observed brightness variation in the light curve is due to the presence of star-spots on the stellar photosphere \citep{Fang2016}. Because of their unstable nature, these star-spots can move across the stellar surface; their amplitudes fluctuate significantly within each rotation period. This variability may contribute to the observed radii of young, low-mass active stars being larger than theoretical predictions \citep{Stauffer2007, Jackson2009}. 

\begin{figure}[hbt!]
	\centering
	\includegraphics[angle=0,width=0.48\textwidth]{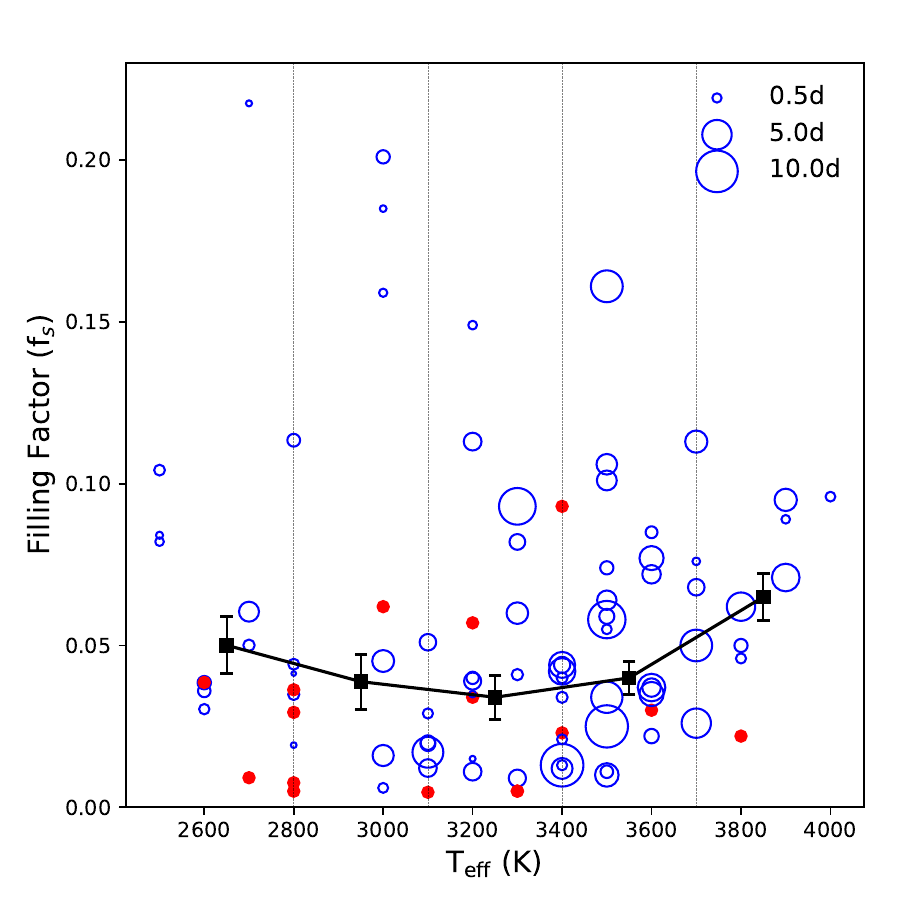}
	\caption{Distribution of derived filling factor ($f_{\mathrm{s}}$) with effective temperature $\teff$. Black square symbols represent the median values within five equally spaced bins. Symbols have the same meaning as in Fig.~\ref{fig4}.}
	\label{fig15}
\end{figure}

\begin{figure}[hbt!]
	\centering
	\includegraphics[angle=0,width=0.48\textwidth]{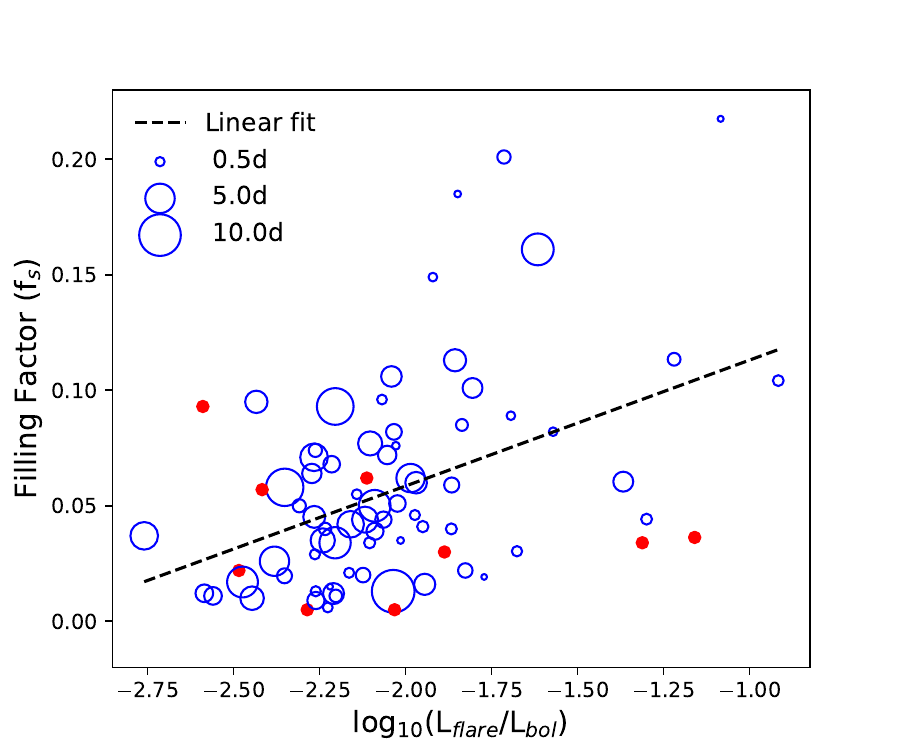}
	\caption{Distribution of derived filling factor (f$_s$) with flare activity ($L_{\rm flare}/L_{\rm bol}$). Black-dashed line shows linear fitting as $f_s = a\,\log_{10}\!\left( \frac{L_{\mathrm{flare}}}{L_{\mathrm{bol}}} \right) + b$, where a=(0.055$\pm$0.013) and b=0.168$\pm$0.027. Symbols have the same meaning as in Fig.~\ref{fig4}.}
	\label{fig16}
\end{figure}

\cite{Kumar2023} using the relation by \citep{Jackson2013} calculated the starspot filling factor ($f_{\mathrm{s}}$), which gives the fractional area covered by starspots (\(A_{\mathrm{spot}} / A_{\star}\)). In figure~\ref {fig15}, we show the correlation between the $f_{\mathrm{s}}$ and $\teff$. We find that the $f_{\mathrm{s}}$ of M dwarfs in our sample is spread over a wide range, indicating potential effects from other parameters on stellar spot coverage. More interestingly, the $f_{\mathrm{s}}$ shows a decreasing trend from cooler M dwarfs to hotter M dwarfs, with a plateau of spot coverage around $\sim$4$\%$ over the $\teff$ interval from $\sim$2900 K to $\sim$3600 K, indicating a saturation-like phase similar to chromospheric and coronal activities observed in fast-rotating stars. In Figure \ref{fig16}, we show the correlation between the $f_{\mathrm{s}}$ and flare activity. The weak positive trend (black dashed line) suggests that higher flare activity is associated with a larger coverage of active regions on M dwarfs. This is expected as strong magnetic activity is often associated with large star-spot coverage on active M dwarfs \citep{Yang2017, Medina2022}

\section{Conclusion}

In this study, we use photometric data from TESS and spectroscopic data from the MFOSC-P spectrograph. We have characterized the stellar activity through FOR, flare energies, star-spot-filling factors, chromospheric activity indicators such as H$\alpha$ emission, and their dependence on stellar mass, spectral type, and rotation period. For a convective M dwarf, the magnitude of the stellar activity is known to be linked to the magnetic properties and state of the star. 

Our study demonstrates that the flaring in M dwarfs is linked to their spectral type, mass, and rotation. For early-to-mid-type M dwarfs (M0–M4), we find that FOR exhibits a nearly flat distribution, which starts declining for later spectral types. This suggests that a transition in magnetic activity occurs near M4, possibly due to the transition to fully convective interiors. Rapid rotators ($P_{\mathrm{rot}}$ < 1 day) exhibit significantly higher FOR, supporting the idea that strong magnetic dynamos in fast-rotating M dwarfs sustain frequent flaring activity.

The negative correlation between FOR and flare amplitude indicates that, more frequently, flaring M dwarfs tend to produce less energetic flares. Our results align with previous findings, which suggested that highly active stars dissipate magnetic energy through numerous low-energy flares rather than fewer high-energy events. Furthermore, our analysis of FFDs across different spectral types reveals that the power-law index ($\alpha$) systematically increases from M0 ($\alpha$ $\sim$ 1.68) to M5 ($\alpha$ $\sim$ 1.95). This trend suggests that flare frequency distributions become steeper for later-type M dwarfs, meaning high-energy flares become even rarer.

Chromospheric activity, traced using the H$\alpha$ equivalent width (EW) and $L_{\mathrm{H}\alpha}/L_{\mathrm{bol}}$, shows that M dwarfs with stronger H$\alpha$ activity tend to exhibit higher FOR. Our spectroscopic analysis reveals a 92$\%$  decrease in $\Delta$EW across the studied stellar mass range, which may be attributed to fundamental changes in the magnetic field generation process as M dwarfs transition to fully convective interiors.

We also find that the stellar spot-filling factor ($f_s$) decreases with increasing temperature and exhibits a saturation-like behavior in stars with $\teff$ $\sim$ 2900 K -- 3600 K. This suggests that strong magnetic activity in fast-rotating M dwarfs results in substantial spot coverage, which could impact their observed stellar radii and rotational evolution.

Our analysis also confirms that there is a strong correlation between chromospheric  ($L_{\rm H\alpha}/L_{\rm bol}$) and flare activity ($L_{\rm flare}/L_{\rm bol}$), following a power-law relationship. We find that even a small increase in chromospheric activity can lead to a significant rise in flare energy. This supports the idea that superflares in M dwarfs do not necessarily require an extra energy-generation mechanism. Instead, they may naturally result from enhanced magnetic activity.

Our study on flaring M dwarfs contributed to the broader understanding of stellar magnetic activity in these fast-rotating cool objects. High-cadence photometric and spectroscopic observations in the future are crucial to refine these relationships further and extend our understanding of activity trends in slowly rotating M dwarfs ($P_{\rm rot} > 10$ days).

\begin{table*}[htbp]
\scriptsize
\centering
\caption{The table describes the total number of sources with TESS (SPOC/TESS-SPOC) data, flaring sources, total detected flares, and $\alpha$ values for the Flare Frequency Distribution (FFD). The percentage of all flares with a duration less than 10 minutes is also shown.}
\begin{tabular}{lcccccc}
\hline
Sp. Type & Total Sources with  & Flaring Sources & No. of Flares & Flares \%  & $\alpha$ (FFD) \\
 & TESS data  &  &  & ($t_{\rm flare} < 10$ min) &  \\
\hline
M0-M0.5 & 8   & 8 & 430 & 74.8 & 1.678$\pm$0.007 \\
M1-M1.5 & 13   & 12 & 583 & 68.5 & 1.705$\pm$0.006 \\
M2-M2.5 & 11   & 10 & 862 & 65.3 & 1.825$\pm$0.008 \\
M3-M3.5 & 18   & 18 & 498 & 65.4 & 1.901$\pm$0.011 \\
M4-M4.5 & 21   & 21 & 1150 & 71.0 & 1.833$\pm$0.003 \\
M5-M5.5 & 16   & 15 & 462 & 77.9 & 1.951$\pm$0.006\\
M6-M6.5 & 9   & 4 & 117 & 85.9 & -- \\
M7-M7.5 & 6   & 5 & 33 & -- & -- \\
M8-M8.5 & 4   & 2 & 3 & -- & -- \\
\hline
\end{tabular}
\label{tab:ffd}
\end{table*}

\section*{Data availability}
Tables \ref{tab:appendix_A1}, \ref{tab:appendix_A2} and \ref{tab:appendix_B} are only available in electronic form at the CDS via anonymous ftp to cdsarc.u-strasbg.fr (130.79.128.5) or via http://cdsweb.u-strasbg.fr/cgi-bin/qcat?J/A+A/.

\begin{acknowledgements}
We thank the referee, Alexander Binks, for his useful comments on the paper. ASR is thankful to the staff at the Mount Abu Observatory for their invaluable assistance during the observations. The MFOSC-P instrument is funded by the Department of Space, Government of India, through the Physical Research Laboratory. MKS thanks the Director, PRL, for supporting the MFOSC-P development program. VK thanks the German Federal Department for Education and Research (Bundesministerium f\"ur Bildung und Forschung - BMBF) under grant agreement (Verbundforschung) number 05A23PK1 for partially supporting this research. L.L. gratefully acknowledges the Collaborative Research Centre 1601 funded by the Deutsche Forschungsgemeinschaft (DFG, German Research Foundation)—SFB 1601 [sub-project A3]— 500700252. J.G.F-T gratefully acknowledges the grant support provided by ANID Fondecyt Postdoc No. 3230001(Sponsoring researcher), and from the Joint Committee ESO-Government of Chile under the agreement 2023 ORP 062/2023. The data from the Transiting Exoplanet Survey Satellite (TESS) was acquired using the tessilator software package. This research has made use of the Exoplanet Follow-up Observation Program (ExoFOP; DOI: 10.26134/ExoFOP5) website, which is operated by the California Institute of Technology, under contract with the National Aeronautics and Space Administration under the Exoplanet Exploration Program.
\end{acknowledgements}

\bibliographystyle{aa}
\bibliography{ref_arvind}

\onecolumn
\begin{appendix}
\section{Summary of TESS observations of targets}

{\scriptsize
\begin{longtable}{lccccccc}
\caption{Summary of TESS observations of flaring targets. Sectors are grouped and years shortened. 
    Pipeline and cadence values include all those used in any observation of each source. 
    Stars where flux contamination could not be calculated (not identified in \textit{Gaia} DR3) 
    are marked with $^{\dagger}$.}\label{tab:appendix_A1}\\
\hline\hline
Source Name & TIC ID & TESS mag & Distance (pc) & Sectors & Years & Pipeline & Cadence \\
\hline
\endfirsthead
\caption[]{Summary of TESS observations of flaring targets (continued).}\\
\hline
Source Name & TIC ID & TESS mag & Distance (pc) & Sectors & Years & Pipeline & Cadence \\
\hline
\endhead
\hline
\endfoot
\hline
\endlastfoot
  PM J03416+5513  &  450182870  &    9.61  &   35.81  &  19,59  &  2019,2022  &  SPOC  &  120  \\
  PM J07151+1555  &  440779601  &   10.15  &   53.30  &  33,44,45,46,71  &  2020,2021,2023  &  SPOC,TESS-SPOC  &  120,200  \\
  PM J23083-1524  &  5656273  &    9.21  &   24.89  &  02,29,42,69  &  2018,2020,2021,2023  &  SPOC  &  120,20  \\
  PM J03322+4914S  &  354790980  &   10.24  &   38.86  &  18  &  2019  &  SPOC  &  120  \\
  PM J04595+0147  &  452763353  &    8.43  &   24.38  &  05,32  &  2018,2020  &  SPOC  &  120  \\
  PM J10143+2104  &  95339414  &   11.47  &   23.38  &  45,46,48  &  2021,2022  &  SPOC  &  20  \\
  PM J19026+3231  &  41840710  &    9.85  &   36.58  &  14,53,54  &  2019,2022  &  SPOC  &  120  \\
  PM J23060+6355  &  435160829  &    9.06  &   24.06  &  17,18,24,57,58  &  2019,2020,2022  &  SPOC  &  120,20  \\
  PM J06310+5002  &  253011416  &    9.22  &   20.66  &  20,60  &  2019,2022  &  SPOC  &  120  \\
  PM J08317+0545  &  265206385  &   10.22  &   47.90  &  07,34,61  &  2019,2021,2023  &  SPOC,TESS-SPOC  &  120,20,600  \\
  PM J09193+6203  &  86232609  &    9.56  &   38.05  &  20,21,47,60  &  2019,2020,2021,2022  &  SPOC  &  120,20  \\
  PM J12576+3513E  &  165916576  &    8.67  &   21.34  &  22,49,76  &  2020,2022,2024  &  SPOC  &  120,20  \\
  PM J15238+5609  &  165720650  &   10.09  &   50.80  &  16,22,23,24,49,50,51,76,77,78  &  2019,2020,2022,2024  &  SPOC  &  120  \\
  PM J15581+4927  &  310170499  &   10.19  &   37.91  &  23,24,50,51  &  2020,2022  &  SPOC  &  120,20  \\
  PM J00428+3532  &  267802440  &    8.51  &   21.71  &  17,57  &  2019,2022  &  SPOC  &  120  \\
  PM J05402+1239  &  127227316  &    9.41  &   34.35  &  06,43,44,45,71  &  2018,2021,2023  &  SPOC,TESS-SPOC  &  120,200  \\
  PM J06262+2349  &  430213846  &    9.95  &   27.46  &  71,72  &  2023  &  SPOC  &  120  \\
  PM J07295+3556  &  18745943  &   10.01  &   42.34  &  20,47,60  &  2019,2021,2022  &  SPOC,TESS-SPOC  &  120,200  \\
  PM J13007+1222  &  88138162  &    7.84  &   11.51  &  23,50  &  2020,2022  &  SPOC  &  120,20  \\
  PM J22387-2037  &  262039241  &    7.06  &    8.90  &  69  &  2023  &  SPOC  &  20  \\
  PM J04284+1741$^\dagger$  &  245791900  &   10.35  &   27.95  &  43,44,70,71  &  2021,2023  &  SPOC,TESS-SPOC  &  20,200  \\
  PM J06212+4414  &  189882802  &   10.29  &   36.92  &  20,60  &  2019,2022  &  SPOC,TESS-SPOC  &  120,200  \\
  PM J11201-1029  &  453465810  &    9.20  &   18.90  &  09,36,63  &  2019,2021,2023  &  SPOC  &  120  \\
  PM J13518+1247  &  72546623  &   10.18  &   26.67  &  23,50  &  2020,2022  &  SPOC  &  120,20  \\
  PM J15218+2058  &  355793860  &    8.02  &   11.44  &  24,51,78  &  2020,2022,2024  &  SPOC  &  120,20  \\
  PM J16170+5516  &  207436278  &    8.08  &   20.25  &  16,23-25,49-52,56,76-79  &  2019,2020,2022,2024  &  SPOC  &  120,20  \\
  PM J09177+4612  &  56914413  &    9.53  &   31.45  &  21  &  2020  &  SPOC  &  120  \\
  PM J10043+5023  &  88723334  &    9.48  &   21.79  &  21,48,75  &  2020,2022,2024  &  SPOC  &  120,20  \\
  PM J11519+0731  &  291074569  &   10.39  &   15.77  &  22,45,46,49  &  2020,2021,2022  &  SPOC  &  120,20  \\
  PM J15557+6840  &  272785770  &   10.01  &   25.63  &  14,16-26,40,41,47-57,60  &  2019,2020,2021,2022  &  SPOC,TESS-SPOC  &  120,20,200  \\
  PM J04333+2359$^\dagger$  &  268397675  &   11.00  &  -  &  43,44,70,71  &  2021,2023  &  SPOC,TESS-SPOC  &  20,200  \\
  PM J05091+1527  &  293769522  &   10.28  &   29.72  &  32,43,71  &  2020,2021,2023  &  SPOC,TESS-SPOC  &  120,200  \\
  PM J05337+0156  &  220044948  &    9.21  &   15.72  &  06,32  &  2018,2020  &  SPOC  &  120,20  \\
  PM J05547+1055  &  139646768  &   10.28  &   24.66  &  33,71  &  2020,2023  &  SPOC,TESS-SPOC  &  120,200  \\
  PM J07319+3613S$^\dagger$  &  16034014  &    8.37  &   12.00  &  20,60  &  2019,2022  &  SPOC  &  120  \\
  PM J07349+1445  &  14768025  &    9.33  &   16.18  &  07,44,45,46,71  &  2019,2021,2023  &  SPOC,TESS-SPOC  &  120,200  \\
  PM J11529+3554  &  144401584  &   11.47  &   39.78  &  49  &  2022  &  SPOC  &  20  \\
  PM J12355+2439  &  376524964  &   11.34  &   75.74  &  49  &  2022  &  SPOC  &  120  \\
  PM J13352+1714  &  95565768  &   11.21  &   73.07  &  23,50  &  2020,2022  &  SPOC  &  120  \\
  PM J14137+4618  &  168704721  &   10.94  &   39.27  &  49,50  &  2022  &  SPOC  &  120  \\
  PM J04238+1455  &  435930877  &   10.91  &   37.97  &  05,32,43,44,70,71  &  2018,2020,2021,2023  &  SPOC,TESS-SPOC  &  120,20,200  \\
  PM J09302+2630  &  172517469  &   10.42  &   24.28  &  21,44,45,46,48  &  2020,2021,2022  &  SPOC  &  120,20  \\
  PM J09557+3521  &  4972033  &   10.41  &   19.22  &  21,48  &  2020,2022  &  SPOC  &  120,20  \\
  PM J12485+4933  &  51269322  &   10.14  &   36.41  &  15,49  &  2019,2022  &  SPOC  &  120,20  \\
  PM J12490+6606  &  341687654  &    8.42  &   12.80  &  15,21,22  &  2019,2020  &  SPOC  &  120  \\
  PM J13417+5815$^\dagger$  &  141816993  &   10.25  &   31.57  &  15,16,48,49  &  2019,2022  &  SPOC  &  120,20  \\
  PM J16591+2058$^\dagger$  &  345727464  &    9.94  &   20.70  &  25  &  2020  &  SPOC  &  120  \\
  PM J00325+0729$^\dagger$  &  468360701  &   10.46  &   35.57  &  42,43,70  &  2021,2023  &  SPOC,TESS-SPOC  &  20,200  \\
  PM J01593+5831  &  445501347  &    9.53  &   13.13  &  58  &  2022  &  SPOC  &  120  \\
  PM J02088+4926  &  250602194  &   10.03  &   17.06  &  18,58  &  2019,2022  &  SPOC  &  120,20  \\
  PM J05062+0439  &  455029978  &   10.66  &   27.78  &  05  &  2018  &  SPOC  &  120  \\
  PM J06000+0242  &  282501711  &    8.58  &    5.21  &  06,33  &  2018,2020  &  SPOC  &  120,20  \\
  PM J07033+3441  &  165204611  &   10.46  &   13.26  &  20,44,45,47,60  &  2019,2021,2022  &  SPOC  &  120,20  \\
  PM J07100+3831  &  321103619  &    8.55  &    6.07  &  20,47  &  2019,2021  &  SPOC  &  120,20  \\
  PM J09161+0153  &  290474796  &   10.41  &   15.64  &  08,61  &  2019,2023  &  SPOC  &  120,20  \\
  PM J10357+0215  &  374239576  &   11.44  &   28.04  &  35,45,46  &  2021  &  SPOC,TESS-SPOC  &  20,600  \\
  PM J10360+0507  &  374267666  &   10.08  &   15.29  &  45,46  &  2021  &  SPOC  &  20  \\
  PM J11033+1337  &  427586730  &   10.37  &   15.24  &  22,45,46,49,72  &  2020,2021,2022,2023  &  SPOC  &  120,20  \\
  PM J11118+3332S  &  85334035  &    9.88  &   13.35  &  22,48  &  2020,2022  &  SPOC  &  120,20  \\
  PM J12156+5239  &  416538839  &   10.14  &   29.52  &  22,48,49  &  2020,2022  &  SPOC  &  120,20  \\
  PM J13536+7737  &  219463771  &   10.25  &   13.25  &  14,15,20,21,26,40,41,47,48,53,60  &  2019,2020,2021,2022  &  SPOC  &  120,20  \\
  PM J15126+4543  &  233644917  &   10.53  &   28.52  &  16,23,24,50,51  &  2019,2020,2022  &  SPOC  &  120  \\
  PM J05243-1601$^\dagger$  &  442932272  &  -  &  -  &  32  &  2020  &  SPOC  &  120  \\
  PM J13317+2916  &  368129164  &    9.29  &   15.59  &  23,50,77  &  2020,2022,2024  &  SPOC  &  120,20  \\
  PM J17199+2630W  &  257798327  &    8.89  &   10.75  &  25,26,52  &  2020,2022  &  SPOC  &  120,20  \\
  PM J01033+6221  &  52183206  &   10.48  &    9.84  &  18,24,58  &  2019,2020,2022  &  SPOC  &  120,20  \\
  PM J02002+1303  &  404715018  &    9.30  &    4.47  &  70,71  &  2023  &  SPOC  &  120  \\
  PM J06579+6219$^\dagger$  &  88384320  &   10.31  &   11.45  &  20,47,60  &  2019,2021,2022  &  SPOC  &  120,20  \\
  PM J07364+0704  &  234479593  &    9.94  &    8.50  &  07,34  &  2019,2021  &  SPOC  &  120,20  \\
  PM J09449-1220  &  289706625  &   10.42  &   13.12  &  08,35,62  &  2019,2021,2023  &  SPOC  &  120,20  \\
  PM J12142+0037  &  397052407  &   10.37  &    8.08  &  46  &  2021  &  SPOC  &  20  \\
  PM J13005+0541  &  411248800  &   10.37  &    8.56  &  23,50  &  2020,2022  &  SPOC  &  120,20  \\
  PM J20298+0941  &  374416052  &    9.97  &    7.47  &  54,55  &  2022  &  SPOC  &  20  \\
  PM J12332+0901$^\dagger$  &  399087412  &    9.00  &  -  &  23,46  &  2020,2021  &  SPOC  &  120,20  \\
  PM J17338+1655  &  400361232  &   10.89  &   16.42  &  26,52,53  &  2020,2022  &  SPOC  &  120  \\
  PM J10564+0700  &  365006789  &    9.28  &    2.41  &  45,46  &  2021  &  SPOC  &  20  \\
                G99-049  &  282501711  &    8.58  &    5.20  &  06,33  &  2018,2020  &  SPOC  &  120,20  \\
                LHS1723  &  43605290  &    9.47  &    5.37  &  05,32  &  2018,2020  &  SPOC  &  120,20  \\
                 L449-1  &  77957301  &    9.05  &   11.71  &  05,06,32  &  2018,2020  &  SPOC  &  120,20  \\
                  GL285  &  266744225  &    8.34  &    5.99  &  07,34  &  2019,2021  &  SPOC  &  120,20  \\
                 GJ1156  &  389356212  &   10.43  &    6.47  &  49  &  2022  &  SPOC  &  20  \\
DENIS-PJ213422.2-431610  &  147207061  &  -  &  -  &  01,28,68  &  2018,2020,2023  &  SPOC  &  120  \\
 2MASSJ00244419-2708242  &  340703996  &   11.33  &    7.73  &  02,29,69  &  2018,2020,2023  &  SPOC  &  120,20  \\
 2MASSJ00045753-1709369  &  289575374  &   13.04  &   17.03  &  02,29,69  &  2018,2020,2023  &  SPOC  &  120  \\
 2MASSJ20021341-5425558  &  201688405  &   13.84  &   18.07  &  13,27,67  &  2019,2020,2023  &  SPOC  &  120,20  \\
               LP731-47  &  1539914  &   14.08  &   23.84  &  09,36  &  2019,2021  &  SPOC  &  120,20  \\
                 GJ3622  &  55099399  &   11.34  &    4.56  &  09,62  &  2019,2023  &  SPOC  &  120,20  \\
 2MASSJ02141251-0357434  &  471012790  &   12.55  &   12.49  &  04,31  &  2018,2020  &  SPOC  &  120,20  \\
 2MASSJ10031918-0105079  &  178835398  &   14.76  &   20.12  &  62,72  &  2023  &  SPOC,TESS-SPOC  &  20,200  \\
 2MASSJ09522188-1924319  &  415678624  &   14.10  &   28.31  &  08,35,62  &  2019,2021,2023  &  SPOC  &  120  \\
 2MASSJ04291842-3123568  &  170675902  &   13.52  &   16.83  &  04,05,31,32  &  2018,2020  &  SPOC  &  120  \\
 2MASSJ23062928-0502285  &  278892590  &   13.85  &   69.57  &  70  &  2023  &  SPOC  &  20  \\
 2MASSJ04351612-1606574  &  117733581  &   12.84  &   10.61  &  05,32  &  2018,2020  &  SPOC  &  120  \\
 2MASSJ22264440-7503425  &  317021872  &   14.99  &   23.45  &  27,67  &  2020,2023  &  SPOC  &  120,20  \\
 2MASSJ23312174-2749500  &  304392811  &   14.08  &   13.63  &  02,29,69  &  2018,2020,2023  &  SPOC  &  120,20  \\

\end{longtable}
}

\onecolumn
{\scriptsize
\begin{longtable}{lccccccc}
\caption{Summary of TESS observations of targets for non-flaring sources.}\label{tab:appendix_A2}\\
\hline
Source Name & TIC ID & TESS mag & Distance (pc) & Sectors & Years & Pipeline & Cadence \\
\hline
\endfirsthead

\caption[]{Summary of TESS observations of targets for non-flaring sources (continued).}\\
\hline
Source Name & TIC ID & TESS mag & Distance (pc) & Sectors & Years & Pipeline & Cadence \\
\hline
\endhead

\hline
\endfoot

\hline
\endlastfoot

PM J15416+1828  &  137054397  & 10.37  & 46.38  & 51        & 2022       & TESS-SPOC        & 600      \\
PM J06596+0545  &  237864235  & 10.36  & 29.13  & 33        & 2020       & SPOC             & 120      \\
2MASSJ02591181+0046468 & 347548361 & 13.51 & 44.52 & 04        & 2018       & SPOC             & 120      \\
LP844-25       &  2092269    & 14.26  & 25.82  & 35,61,62  & 2021,2023  & SPOC             & 120,20   \\
2MASSJ21322975-0511585 & 268958223 & 13.53 & 20.09 & 55        & 2022       & SPOC             & 120      \\
2MASSWJ1012065-304926  & 71028960  & 14.57  & 22.65  & 35,36,62,63 & 2021,2023 & SPOC          & 120,20   \\
2MASSJ23155449-0627462 & 4609881   & 13.22  & 16.82  & 42,70     & 2021,2023  & SPOC,TESS-SPOC  & 120,200  \\
2MASSJ05023867-3227500 & 1527678   & 14.61  & 29.29  & 05,32     & 2018,2020  & SPOC            & 120      \\
2MASSJ03313025-3042383 & 142944290 & 13.84  & 12.51  & 04,31     & 2018,2020  & SPOC            & 120      \\
2MASSJ02484100-1651216 & 29959288  & 15.09  & 22.43  & 04,31     & 2018,2020  & SPOC            & 120      \\
2MASSJ03061159-3647528 & 308243298 & 14.99  & 13.26  & 03,04,30,31 & 2018,2020 & SPOC          & 120      \\
\hline
\end{longtable}
}

\onecolumn
\section{Stellar properties and derived parameters}
{\scriptsize
\begin{longtable}{lcccccccccccc}
\caption{Stellar properties and derived parameters. Spectral type, H$\alpha$ EW, 
    $L_{\rm H\alpha}/L_{\rm bol}$, $T_{\rm eff}$, $P_{\rm rot}$, and $f_{\rm s}$ are 
    from \citet{Kumar2023}. FOR, $L_{\rm flare}/L_{\rm bol}$, $\langle E_{\mathrm{flare}} \rangle$, 
    and $A_{\rm Med}$ are derived in this work (only for sources with TESS data). 
    Stellar mass and radius are from \citet{Stassun2019}.}\label{tab:appendix_B}\\
\hline
Source Name & Spectral & Median & Mean & $T_{\mathrm{eff}}$ & $P_{\mathrm{rot}}$ & $f_{\mathrm{s}}$ & Mass & Radius & FOR & $L_{\mathrm{flare}}/L_{\mathrm{bol}}$ & $\langle E_{\mathrm{flare}} \rangle$ & $A_{\mathrm{Med}}$ \\
 & Type & H$\alpha$ EW (\AA) & $\log_{10}(L_{\mathrm{H}\alpha}/L_{\mathrm{bol}})$ & (K) & (days) & (\%) & ($M_\odot$) & ($R_\odot$) & (\%) & ($\times 10^{-3}$) & ($\times 10^{32}$ erg) & (\%) \\
\hline
\endfirsthead

\caption[]{Stellar properties and derived parameters (continued).}\\
\hline
Source Name & Spectral & Median & Mean & $T_{\mathrm{eff}}$ & $P_{\mathrm{rot}}$ & $f_{\mathrm{s}}$ & Mass & Radius & FOR & $L_{\mathrm{flare}}/L_{\mathrm{bol}}$ & $\langle E_{\mathrm{flare}} \rangle$ & $A_{\mathrm{Med}}$ \\
 & Type & H$\alpha$ EW (\AA) & $\log_{10}(L_{\mathrm{H}\alpha}/L_{\mathrm{bol}})$ & (K) & (days) & (\%) & ($M_\odot$) & ($R_\odot$) & (\%) & ($\times 10^{-3}$) & ($\times 10^{32}$ erg) & (\%) \\
\hline
\endhead

\hline
\endfoot

\hline
\endlastfoot
  PM J03416+5513  &    0.0  &   -1.7 $\pm$   0.0  &  -3.984  &   3800  &  4.641  &    6.2  &  0.636  &  0.663  &   0.804  &    10.347  &  49.279  &    0.0214 \\ 
  PM J07151+1555  &    0.0  &   -2.1 $\pm$   0.1  &  -3.854  &   4000  &  0.554  &    9.6  &  0.640  &  0.755  &   0.494  &     8.545  &  31.144  &    0.0117 \\ 
  PM J23083-1524  &    0.0  &   -1.7 $\pm$   0.1  &  -3.980  &   3900  &  0.432  &    8.9  &  0.578  &  0.587  &   0.672  &    20.249  &  29.575  &    0.0131 \\ 
  PM J03322+4914S  &    0.5  &   -1.6 $\pm$   0.1  &  -3.960  &   3700  &  5.799  &    5.0  &  0.573  &  0.581  &   0.370  &     8.136  &  14.393  &    0.0167 \\ 
  PM J04595+0147  &    0.5  &   -1.5 $\pm$   0.0  &  -4.033  &   3900  &  4.406  &    7.1  &  0.600  &  0.868  &   1.077  &     5.423  &  60.474  &    0.0066 \\ 
  PM J10143+2104  &    0.5  &   -1.8 $\pm$   0.0  &  -3.973  &   3800  &  -  &  -  &  0.677  &  0.719  &   0.584  &    14.836  &  13.828  &    0.0073 \\ 
  PM J19026+3231  &    0.5  &   -2.5 $\pm$   0.1  &  -3.815  &   3700  &  0.347  &    7.6  &  0.605  &  0.621  &   0.629  &     9.376  &  30.547  &    0.0097 \\ 
  PM J23060+6355  &    0.5  &   -1.6 $\pm$   0.0  &  -4.057  &   3700  &  2.848  &   11.3  &  0.300  &  0.599  &   0.318  &    13.932  &  17.080  &    0.0147 \\ 
  PM J06310+5002  &    1.0  &   -2.1 $\pm$   0.1  &  -3.907  &   3700  &  4.966  &    2.6  &  0.523  &  0.526  &   0.575  &     4.161  &   7.221  &    0.0098 \\ 
  PM J08317+0545  &    1.0  &   -2.9 $\pm$   0.1  &  -3.781  &   3800  &  0.595  &    4.6  &  0.644  &  0.673  &   0.391  &    10.650  &  30.538  &    0.0186 \\ 
  PM J09193+6203  &    1.0  &   -2.2 $\pm$   0.1  &  -3.865  &   3600  &  -  &    3.0  &  0.590  &  0.813  &   3.619  &    12.981  &  163.412  &    0.0105 \\ 
  PM J12576+3513E  &    1.0  &   -2.0 $\pm$   0.1  &  -3.928  &   3600  &  3.355  &    7.7  &  0.578  &  0.588  &   0.311  &     7.894  &   6.414  &    0.0127 \\ 
  PM J15238+5609  &    1.0  &   -1.7 $\pm$   0.1  &  -3.994  &   3800  &  1.003  &    5.0  &  0.610  &  0.802  &   0.471  &     4.913  &  25.693  &    0.0133 \\ 
  PM J15581+4927  &    1.0  &   -2.6 $\pm$   0.1  &  -3.897  &   3600  &  0.815  &    8.5  &  0.600  &  0.615  &   0.371  &    14.593  &  24.517  &    0.0272 \\ 
  PM J00428+3532  &    1.5  &   -2.5 $\pm$   0.1  &  -3.878  &   3500  &  2.174  &    6.4  &  0.649  &  0.680  &   1.276  &     5.344  &  11.896  &    0.0066 \\ 
  PM J05402+1239  &    1.5  &   -2.2 $\pm$   0.1  &  -3.931  &   3700  &  1.589  &    6.8  &  0.666  &  0.704  &   0.582  &     6.102  &  16.808  &    0.0094 \\ 
  PM J06262+2349  &    1.5  &   -1.5 $\pm$   0.1  &  -4.039  &   3500  &  8.059  &    5.8  &  0.512  &  0.515  &   0.254  &     4.456  &   5.065  &    0.0107 \\ 
  PM J07295+3556  &    1.5  &   -2.9 $\pm$   0.1  &  -3.775  &   3600  &  1.979  &    7.2  &  0.646  &  0.676  &   0.540  &     8.847  &  21.633  &    0.0179 \\ 
  PM J13007+1222  &    1.5  &   -2.1 $\pm$   0.1  &  -3.986  &   3900  &  2.881  &    9.5  &  0.556  &  0.563  &   0.900  &     3.685  &   5.028  &    0.0055 \\ 
  PM J15416+1828  &    1.5  &   -2.2 $\pm$   0.1  &  -3.965  &   3600  &  -  &  -  &  0.497  &  0.467  &  -  &  -  &   -  &  -\\ 
  PM J22387-2037  &    1.5  &   -2.3 $\pm$   0.1  &  -3.869  &   3600  &  4.391  &    3.7  &  0.589  &  0.602  &   1.940  &     1.740  &   1.259  &    0.0058 \\ 
  PM J04284+1741  &    2.0  &   -3.0 $\pm$   0.1  &  -3.859  &   3500  &  2.449  &   10.6  &  0.391  &  0.356  &   0.392  &     9.103  &   4.849  &    0.0232 \\ 
  PM J06212+4414  &    2.0  &   -2.9 $\pm$   0.1  &  -3.841  &   3500  &  5.848  &   16.1  &  0.540  &  0.623  &   0.395  &    24.243  &  44.842  &    0.0278 \\ 
  PM J11201-1029  &    2.0  &   -2.0 $\pm$   0.1  &  -3.976  &   3500  &  5.652  &    3.4  &  0.507  &  0.510  &   0.269  &     6.242  &   8.242  &    0.0115 \\ 
  PM J13518+1247  &    2.0  &   -2.1 $\pm$   0.1  &  -3.997  &   3800  &  -  &    2.2  &  0.466  &  0.469  &   0.619  &     3.280  &   7.672  &    0.0205 \\ 
  PM J15218+2058  &    2.0  &   -2.6 $\pm$   0.1  &  -3.866  &   3600  &  3.395  &    3.5  &  0.527  &  0.530  &   1.483  &     5.748  &   6.663  &    0.0080 \\ 
  PM J16170+5516  &    2.0  &   -2.5 $\pm$   0.1  &  -3.900  &   3600  &  1.975  &    3.7  &  -  &  -  &   1.132  &  -  &   -  &    0.0082 \\ 
  PM J06596+0545  &    2.5  &   -1.8 $\pm$   0.1  &  -4.062  &   3500  &  -  &  -  &  0.483  &  0.485  &  -  &  -  &   -  &  -\\ 
  PM J09177+4612  &    2.5  &   -3.9 $\pm$   0.1  &  -3.693  &   3500  &  1.016  &    7.4  &  0.651  &  0.683  &   0.681  &     5.471  &   8.904  &    0.0121 \\ 
  PM J10043+5023  &    2.5  &   -3.6 $\pm$   0.1  &  -3.739  &   3500  &  1.315  &    5.9  &  0.520  &  0.522  &   0.513  &    13.625  &  12.378  &    0.0185 \\ 
  PM J11519+0731  &    2.5  &   -3.0 $\pm$   0.1  &  -3.858  &   3500  &  2.283  &   10.1  &  0.291  &  0.278  &   0.230  &    15.663  &   3.017  &    0.0227 \\ 
  PM J15557+6840  &    2.5  &   -2.4 $\pm$   0.1  &  -3.907  &   3400  &  3.917  &    4.4  &  0.488  &  0.490  &   0.286  &     7.628  &   6.130  &    0.0238 \\ 
  PM J04333+2359  &    3.0  &   -3.2 $\pm$   0.1  &  -3.846  &   3500  &  -  &  -  &  -  &  -  &   0.241  &  -  &   -  &    0.0354 \\ 
  PM J05091+1527  &    3.0  &   -3.2 $\pm$   0.1  &  -3.829  &   3400  &  2.532  &    1.2  &  0.506  &  0.508  &   0.351  &     6.178  &   5.592  &    0.0179 \\ 
  PM J05337+0156  &    3.0  &   -6.4 $\pm$   0.1  &  -3.510  &   3400  &  0.605  &    1.3  &  0.461  &  0.464  &   1.476  &     5.485  &   3.071  &    0.0146 \\ 
  PM J05547+1055  &    3.0  &   -4.0 $\pm$   0.1  &  -3.739  &   3500  &  0.565  &    5.5  &  0.439  &  0.443  &   0.426  &     7.219  &   7.084  &    0.0192 \\ 
  PM J07319+3613S  &    3.0  &   -1.9 $\pm$   0.1  &  -4.136  &   3400  &  -  &    9.3  &  0.516  &  0.518  &   0.583  &     2.576  &   2.863  &    0.0084 \\ 
  PM J07349+1445  &    3.0  &   -2.5 $\pm$   0.1  &  -3.967  &   3400  &  -  &    2.3  &  -  &  -  &   0.603  &  -  &   -  &    0.0060 \\ 
  PM J11529+3554  &    3.0  &   -4.7 $\pm$   0.2  &  -3.628  &   3300  &  2.690  &    6.0  &  0.419  &  0.425  &   0.070  &    10.737  &   1.714  &    0.0441 \\ 
  PM J12355+2439  &    3.0  &   -2.5 $\pm$   0.1  &  -3.913  &   3500  &  -  &  -  &  0.646  &  0.676  &   0.259  &     9.557  &  47.451  &    0.0599 \\ 
  PM J13352+1714  &    3.0  &   -2.7 $\pm$   0.1  &  -  &   3600  &  1.213  &    2.2  &  0.634  &  0.659  &   0.204  &    14.924  &  35.993  &    0.0227 \\ 
  PM J14137+4618  &    3.0  &   -4.5 $\pm$   0.1  &  -3.768  &   3400  &  1.536  &    4.4  &  0.502  &  0.504  &   1.156  &     8.628  &  16.897  &    0.0347 \\ 
  PM J04238+1455  &    3.5  &   -5.1 $\pm$   0.4  &  -3.711  &   3500  &  0.923  &    1.1  &  0.542  &  0.546  &   0.267  &     6.273  &   7.356  &    0.0328 \\ 
  PM J09302+2630  &    3.5  &   -3.1 $\pm$   0.1  &  -3.893  &   3400  &  10.517  &    1.3  &  0.425  &  0.430  &   0.126  &     9.212  &   3.059  &    0.0266 \\ 
  PM J09557+3521  &    3.5  &   -3.0 $\pm$   0.1  &  -3.915  &   3300  &  -  &    0.5  &  0.342  &  0.357  &   0.440  &     5.181  &   2.127  &    0.0237 \\ 
  PM J12485+4933  &    3.5  &   -5.2 $\pm$   0.1  &  -3.641  &   3400  &  0.588  &    2.1  &  0.468  &  0.471  &   1.257  &     6.859  &   7.628  &    0.0265 \\ 
  PM J12490+6606  &    3.5  &   -1.0 $\pm$   0.1  &  -4.387  &   3500  &  3.197  &    1.0  &  0.437  &  0.441  &   0.128  &     3.582  &   2.710  &    0.0187 \\ 
  PM J13417+5815  &    3.5  &   -2.3 $\pm$   0.1  &  -4.008  &   3300  &  1.709  &    0.9  &  0.379  &  0.389  &   0.202  &     5.491  &   2.612  &    0.0254 \\ 
  PM J16591+2058  &    3.5  &   -3.2 $\pm$   0.1  &  -3.880  &   3400  &  4.075  &    4.2  &  0.398  &  0.406  &   0.587  &     6.928  &   4.212  &    0.0200 \\ 
  PM J00325+0729  &    4.0  &   -6.0 $\pm$   0.1  &  -3.629  &   3200  &  1.820  &   11.3  &  -  &  -  &   0.163  &  -  &   -  &    0.0283 \\ 
  PM J01593+5831  &    4.0  &   -6.1 $\pm$   0.1  &  -3.700  &   3200  &  -  &  -  &  0.381  &  0.391  &   1.078  &     4.884  &   1.620  &    0.0134 \\ 
  PM J02088+4926  &    4.0  &   -7.2 $\pm$   0.2  &  -3.534  &   3300  &  0.749  &    4.1  &  0.373  &  0.383  &   1.562  &    11.229  &   7.356  &    0.0228 \\ 
  PM J05062+0439  &    4.0  &   -7.3 $\pm$   0.2  &  -3.662  &   3200  &  0.889  &    4.0  &  0.465  &  0.468  &   0.465  &     5.841  &   3.693  &    0.0258 \\ 
  PM J06000+0242  &    4.0  &   -2.8 $\pm$   0.1  &  -4.042  &   3200  &  1.807  &    1.1  &  0.232  &  0.261  &   0.416  &     2.758  &   0.320  &    0.0137 \\ 
  PM J07033+3441  &    4.0  &   -5.1 $\pm$   0.1  &  -3.730  &   3200  &  -  &    3.4  &  0.251  &  0.277  &   0.539  &    48.805  &  11.108  &    0.0292 \\ 
  PM J07100+3831  &    4.0  &   -1.9 $\pm$   0.1  &  -4.296  &   3100  &  5.483  &    1.7  &  0.298  &  0.319  &   0.089  &     3.359  &   0.482  &    0.0145 \\ 
  PM J09161+0153  &    4.0  &   -4.7 $\pm$   0.1  &  -3.751  &   3300  &  1.432  &    8.2  &  0.290  &  0.312  &   0.537  &     9.256  &   2.308  &    0.0238 \\ 
  PM J10357+0215  &    4.0  &   -3.4 $\pm$   0.2  &  -3.791  &   3400  &  0.707  &    4.0  &  0.320  &  0.299  &   0.207  &    13.593  &   4.536  &    0.0553 \\ 
  PM J10360+0507  &    4.0  &   -5.6 $\pm$   0.1  &  -3.670  &   3300  &  -  &  -  &  0.335  &  0.350  &   0.617  &    17.694  &   4.030  &    0.0261 \\ 
  PM J11033+1337  &    4.0  &   -3.1 $\pm$   0.1  &  -3.909  &   3300  &  -  &    0.5  &  0.289  &  0.311  &   0.430  &     9.298  &   2.544  &    0.0207 \\ 
  PM J11118+3332S  &    4.0  &   -5.4 $\pm$   0.1  &  -3.682  &   3300  &  7.763  &    9.3  &  0.307  &  0.326  &   0.433  &     6.249  &   1.086  &    0.0209 \\ 
  PM J12156+5239  &    4.0  &   -5.4 $\pm$   0.1  &  -3.731  &   3400  &  0.726  &    3.4  &  0.483  &  0.485  &   0.562  &     7.867  &   4.468  &    0.0228 \\ 
  PM J13536+7737  &    4.0  &   -3.5 $\pm$   0.1  &  -3.926  &   3100  &  1.231  &    2.0  &  0.265  &  0.290  &   0.207  &     7.528  &   1.135  &    0.0275 \\ 
  PM J15126+4543  &    4.0  &   -4.0 $\pm$   0.2  &  -3.840  &   3200  &  1.687  &    3.9  &  0.311  &  0.330  &   0.606  &     8.154  &   4.775  &    0.0264 \\ 
  PM J05243-1601  &    4.5  &  -10.5 $\pm$   0.3  &  -3.502  &   3000  &  0.396  &   15.9  &  -  &  -  &   2.910  &  -  &   -  &    0.0123 \\ 
  PM J13317+2916  &    4.5  &   -9.1 $\pm$   0.2  &  -3.565  &   3200  &  0.268  &    3.5  &  0.535  &  0.539  &   3.485  &     9.678  &  10.625  &    0.0079 \\ 
  PM J17199+2630W  &    4.5  &   -2.2 $\pm$   0.1  &  -4.188  &   3200  &  -  &    5.7  &  0.398  &  0.406  &   4.949  &     3.832  &  10.388  &    0.0100 \\ 
  PM J01033+6221  &    5.0  &  -14.4 $\pm$   0.3  &  -3.424  &   3000  &  1.027  &   20.1  &  0.202  &  0.232  &   1.300  &    19.335  &   2.385  &    0.0358 \\ 
  PM J02002+1303  &    5.0  &   -2.1 $\pm$   0.1  &  -4.198  &   3100  &  -  &  -  &  0.150  &  0.180  &   0.545  &     6.207  &   2.149  &    0.0135 \\ 
  PM J06579+6219  &    5.0  &   -2.6 $\pm$   0.1  &  -4.180  &   3000  &  2.587  &    1.6  &  0.239  &  0.267  &   0.232  &    11.372  &   0.975  &    0.0424 \\ 
  PM J07364+0704  &    5.0  &   -5.6 $\pm$   0.2  &  -3.890  &   3000  &  0.571  &    0.6  &  0.216  &  0.245  &   0.609  &     5.943  &   0.615  &    0.0263 \\ 
  PM J09449-1220  &    5.0  &  -13.0 $\pm$   0.3  &  -3.589  &   3200  &  0.442  &   14.9  &  0.287  &  0.309  &   0.891  &    12.012  &   2.463  &    0.0258 \\ 
  PM J12142+0037  &    5.0  &   -7.3 $\pm$   0.2  &  -3.808  &   3100  &  1.584  &    5.1  &  0.179  &  0.210  &   0.176  &     9.475  &   0.376  &    0.0329 \\ 
  PM J13005+0541  &    5.0  &   -7.6 $\pm$   0.2  &  -3.646  &   3100  &  0.600  &    2.9  &  0.180  &  0.211  &   0.234  &     5.454  &   0.232  &    0.0271 \\ 
  PM J20298+0941  &    5.0  &   -5.6 $\pm$   0.2  &  -3.813  &   3000  &  -  &  -  &  0.184  &  0.215  &   0.153  &     6.349  &   0.175  &    0.0327 \\ 
  PM J12332+0901  &    5.5  &   -6.7 $\pm$   0.2  &  -3.872  &   3200  &  0.207  &    1.5  &  0.195  &  0.225  &   0.991  &     6.035  &   0.471  &    0.0144 \\ 
  PM J17338+1655  &    5.5  &  -16.0 $\pm$   0.4  &  -3.526  &   3000  &  0.266  &   18.5  &  0.293  &  0.314  &   0.340  &    14.180  &   2.683  &    0.0550 \\ 
  PM J10564+0700  &    6.0  &   -9.6 $\pm$   0.3  &  -3.746  &   3000  &  -  &    6.2  &  0.109  &  0.135  &   1.068  &     7.719  &   0.164  &    0.0159 \\ 
                G99-049  &    3.5  &   -7.1 $\pm$   0.1  &  -3.229  &   3100  &  1.805  &    1.2  &  0.258  &  0.261  &   0.416  &     2.602  &   0.267  &    0.0137 \\ 
                LHS1723  &    4.0  &   -1.2 $\pm$   0.2  &  -4.155  &   3100  &  -  &    0.5  &  -  &  -  &   0.210  &  -  &   -  &    0.0327 \\ 
                 L449-1  &    4.0  &   -6.8 $\pm$   0.2  &  -3.357  &   3100  &  1.296  &    2.0  &  0.402  &  0.410  &   1.362  &     4.450  &   1.572  &    0.0121 \\ 
                  GL285  &    4.5  &   -9.3 $\pm$   0.1  &  -3.179  &   3000  &  2.770  &    4.5  &  0.315  &  0.333  &   1.969  &     5.429  &   0.842  &    0.0120 \\ 
                 GJ1156  &    5.0  &   -6.9 $\pm$   0.3  &  -3.705  &   2900  &  -  &  -  &  0.142  &  0.172  &   0.521  &    12.032  &   0.305  &    0.0354 \\ 
DENIS-PJ213422.2-431610  &    5.5  &   -0.9 $\pm$   0.5  &  -4.466  &   2800  &  -  &    0.5  &  -  &  -  &   0.017  &  -  &   -  &    0.0995 \\ 
 2MASSJ02591181+0046468  &    5.5  &  -12.0 $\pm$   0.8  &  -3.387  &   2800  &  -  &  -  &  0.215  &  0.245  &  -  &  -  &   -  &  -\\ 
 2MASSJ00244419-2708242  &    5.5  &   -4.0 $\pm$   4.5  &  -3.734  &   2800  &  0.945  &   11.3  &  0.127  &  0.155  &   0.217  &    60.379  &   1.991  &    0.0900 \\ 
 2MASSJ00045753-1709369  &    5.5  &   -4.4 $\pm$   0.4  &  -3.813  &   2800  &  0.192  &    1.9  &  0.121  &  0.148  &   0.030  &    16.930  &   0.440  &    0.0892 \\ 
 2MASSJ20021341-5425558  &    5.5  &   -1.9 $\pm$   0.4  &  -3.785  &   2800  &  0.692  &    4.4  &  0.104  &  0.129  &   0.056  &    50.217  &   1.057  &    0.3765 \\ 
               LP844-25  &    6.0  &   -6.7 $\pm$   1.2  &  -4.610  &   2800  &  -  &  -  &  0.099  &  0.123  &  -  &  -  &   -  &  -\\ 
 2MASSJ21322975-0511585  &    6.0  &   -3.5 $\pm$   0.8  &  -5.057  &   2800  &  -  &  -  &  0.124  &  0.151  &  -  &  -  &   -  &  -\\ 
  2MASSWJ1012065-304926  &    6.0  &  -12.5 $\pm$   0.7  &  -3.904  &   2800  &  -  &  -  &  0.102  &  0.127  &  -  &  -  &   -  &  -\\ 
               LP731-47  &    6.0  &   -0.6 $\pm$   0.7  &  -3.847  &   2800  &  -  &    3.6  &  0.096  &  0.128  &   0.025  &    69.286  &   1.155  &    0.3780 \\ 
 2MASSJ23155449-0627462  &    6.0  &   -5.0 $\pm$   0.4  &  -4.122  &   2800  &  -  &  -  &  0.113  &  0.140  &  -  &  -  &   -  &  -\\ 
                 GJ3622  &    6.5  &   -7.7 $\pm$   0.8  &  -4.393  &   2700  &  -  &    0.9  &  -  &  -  &   0.178  &  -  &   -  &    0.0722 \\ 
 2MASSJ05023867-3227500  &    6.5  &   -3.3 $\pm$   0.7  &  -4.050  &   2700  &  -  &  -  &  0.113  &  0.139  &  -  &  -  &   -  &  -\\ 
 2MASSJ02141251-0357434  &    6.5  &   -7.2 $\pm$   0.4  &  -4.012  &   2700  &  2.296  &    6.0  &  0.106  &  0.141  &   0.047  &    42.958  &   0.389  &    0.5685 \\ 
 2MASSJ10031918-0105079  &    7.0  &  -10.6 $\pm$   1.2  &  -4.201  &   2700  &  0.213  &   21.8  &  0.093  &  0.117  &   0.027  &    82.433  &   0.361  &    0.6679 \\ 
 2MASSJ09522188-1924319  &    7.5  &   -4.5 $\pm$   2.1  &  -3.940  &   2600  &  0.909  &  -  &  0.140  &  0.169  &   0.159  &    49.763  &   1.773  &    0.4036 \\ 
 2MASSJ04291842-3123568  &    7.5  &  -10.9 $\pm$   0.7  &  -3.933  &   2600  &  0.907  &    3.6  &  -  &  -  &   0.104  &  -  &   -  &    0.1543 \\ 
 2MASSJ23062928-0502285  &    7.5  &  -12.6 $\pm$   0.5  &  -4.379  &   2600  &  -  &  -  &  0.081  &  0.102  &   0.214  &     1.758  &   0.020  &    0.2078 \\ 
 2MASSJ03313025-3042383  &    7.5  &   -3.5 $\pm$   0.9  &  -4.068  &   2600  &  -  &  -  &  0.092  &  0.116  &  -  &  -  &   -  &  -\\ 
 2MASSJ04351612-1606574  &    7.5  &   -7.8 $\pm$   1.0  &  -4.206  &   2600  &  0.622  &    3.0  &  0.108  &  0.134  &   0.104  &    21.092  &   0.428  &    0.1986 \\ 
 2MASSJ02484100-1651216  &    8.0  &   -6.5 $\pm$   1.2  &  -4.538  &   2600  &  -  &  -  &  0.095  &  0.119  &  -  &  -  &   -  &  -\\ 
 2MASSJ22264440-7503425  &    8.5  &   -0.6 $\pm$   2.4  &  -4.538  &   2500  &  0.678  &   10.4  &  0.102  &  0.127  &   0.006  &   121.256  &   0.265  &    0.9905 \\ 
 2MASSJ03061159-3647528  &    8.5  &   -8.3 $\pm$   0.7  &  -4.421  &   2500  &  -  &  -  &  0.276  &  0.286  &  -  &  -  &   -  &  -\\ 
 2MASSJ23312174-2749500  &    8.5  &   -5.5 $\pm$   1.0  &  -4.594  &   2500  &  0.431  &    8.2  &  0.088  &  0.112  &   0.017  &    26.861  &   0.159  &    0.4997 \\ 
\hline
\end{longtable}
}
\end{appendix}

\end{document}